\documentclass{emulateapj}
\usepackage{apjfonts}

\begin{document}
\title{The Blazar Sequence and the Cosmic Gamma-Ray Background Radiation
 \\ in the Fermi Era}
\author{Yoshiyuki Inoue and Tomonori Totani}
\affil{Department of Astronomy, Kyoto University, 
Kitashirakawa, Sakyo-ku, Kyoto 606-8502, Japan}
\email{yinoue@kusastro.kyoto-u.ac.jp}

\begin{abstract} 
  We present a new model of the blazar gamma-ray luminosity function
  (GLF) and the spectrum of the extragalactic gamma-ray background
  (EGRB), which is consistent with the observed distributions of EGRET
  blazars.  The unified sequence of blazar spectral energy
  distribution (SED) is taken into account to make a non-trivial
  prediction for the EGRB spectrum and more realistic comparison with
  the data than previous studies. We then try to explain the EGRB data
  by the two AGN populations: one is blazars, and the other is
  non-blazar AGNs that are responsible for the EGRB in the MeV band.
  We find that $\sim$80\% of the EGRB photon flux at $>$ 100 MeV can
  be explained by the sum of the two populations, while $\sim$45 \%
  can be accounted for only by blazars.  The predicted EGRB spectrum
  is in agreement with a wide range of the observed data from X-ray to
  GeV, within the systematic uncertainties in the EGRB determination
  by EGRET.  These results indicate that AGNs including blazars are
  the primary source of EGRB.  Blazars are dominant in EGRB at higher
  energy bands of $\gtrsim$ 100 MeV, while non-blazar AGNs dominate at
  $\lesssim$ 100 MeV.  Almost all of the EGRB flux from blazars will
  be resolved into discrete sources by the {\it Fermi} Gamma-ray Space
  Telescope, while that from non-blazar AGNs will largely remain
  unresolved. Therefore, comparison between the integrated source
  counts and diffuse EGRB flux as a function of photon energy will
  give a simple and clear test of our model. Various quantitative
  predictions for {\it Fermi} observations are also made. Especially,
  our model predicts 600--1200 blazars in all sky down to $2 \times
  10^{-9} \ \rm photons \ cm^{-2} s^{-1}$ ($>$100 MeV), which is
  considerably smaller than most of previous studies.  We find that
  the fraction of EGRB energy flux absorbed in intergalactic medium
  (IGM) is not large, and the cascade component reprocessed in IGM
  does not significantly alter the EGRB spectrum.
\end{abstract}


\section{Introduction}
\label{intro}
The origin of the extragalactic diffuse gamma-ray background (EGRB)
has been discussed in astrophysics for a long time. EGRB 
was first discovered by the \textit{SAS 2} satellite ({Fichtel},
{Simpson}, \& {Thompson} 1978; {Thompson} \& {Fichtel}
1982). Subsequently, the EGRB spectrum was confirmed up to $\sim$ 50
GeV by EGRET (Energetic Gamma-Ray Experiment Telescope) on board the
Compton Gamma Ray Observatory. The EGRB flux is about $1 \times$
10$^{-5}$ photons cm$^{-2}$ s$^{-1}$ sr$^{-1}$ above 100 MeV and the
spectrum is approximately a power law with a photon index of $\sim -2$ in a
wide range of $\sim$30 MeV -- 100 GeV ({Sreekumar} {et~al.}  1998;
{Strong}, {Moskalenko}, \& {Reimer} 2004a).  It should be noted that,
however, measurement of EGRB is not an easy task. The Galactic diffuse
background from cosmic-ray interaction in the Galactic disk is a
strong {\it foreground} emission and must be subtracted to estimate
EGRB. Therefore modeling of this foreground component could induce
significant systematic uncertainties in EGRB measurements ({Keshet},
{Waxman}, \& {Loeb} 2004; {Strong}, {Moskalenko}, \& {Reimer} 2004a,b;
{Kamae}, {Abe}, \& {Koi} 2005; {Kamae} {et~al.}  2006).  A possible
systematic error in the calibration of the EGRET detector may also
have affected the EGRB determination (Stecker et al. 2008).

Although several sources of gamma-rays (e.g., clusters of galaxies or
dark matter annihilation) have been proposed as a significant
contributor to EGRB [see, e.g., {Narumoto} \& {Totani} (2006) and
references therein], active galactic nuclei (AGNs) of the blazar class
have been thought as the primary candidate for the origin of EGRB,
since almost all of the extragalactic gamma-ray sources detected by
EGRET are blazars. The blazar contribution to EGRB has been estimated
by a number of papers ({Padovani} {et~al.}  1993; {Stecker},
{Salamon}, \& {Malkan} 1993; {Salamon} \& {Stecker} 1994; {Chiang}
{et~al.} 1995; {Stecker} \& {Salamon} 1996; {Chiang} \& {Mukherjee}
1998; Mukherjee \& Chiang 1999; {M{\"u}cke} \& {Pohl} 2000; {Narumoto}
\& {Totani} 2006; {Giommi} {et~al.} 2006; {Dermer} 2007; {Pavlidou} \&
{Venters} 2008; {Kneiske} \& {Mannheim} 2008; Bhattacharya et
al. 2008). These previous studies have derived different results by
different approaches about the contribution of blazars to EGRB,
ranging from 20\% to 100 \%. A brief summary of these studies and
comparison with our new result will be presented in
\S\ref{subsection:past}. In our previous study (Narumoto \& Totani 2006,
hereafter NT06), we constructed a gamma-ray luminosity function (GLF)
model based on the picture of luminosity-dependent density evolution
(LDDE), which has been known to describe well the evolution of X-ray
luminosity function (XLF) of AGNs. NT06 found that the LDDE model fits
to the EGRET blazar data better than the pure-luminosity-evolution
(PLE) models employed in most of the previous studies.  It is then
found that only 25--50\% of EGRB can be explained by blazars, with the
GLF model parameters that are consistent with the EGRET blazar data
(flux and redshift distributions).

In most of past studies including NT06, however, blazar spectral
energy distributions (SEDs) were assumed to be a single or broken
power-law for all blazars.  In such a modeling, the predicted EGRB
spectrum is almost obviously determined by the assumed power-law
indices. Blazar SED has been theoretically and observationally studied
in detail ({Ghisellini} \& {Tavecchio} 2008b, and references
therein). From the theoretical point of view, blazar emission is
widely believed to be the sum of the synchrotron (radio to UV bands)
and inverse Compton (dominant in gamma-ray bands) components produced
by the same nonthermal electron population accelerated in relativistic
jets.  The source of target photons for the inverse-Compton (IC)
component could be either the synchrotron photons produced in the jet
itself (synchrotron-self-Compton, SSC), or photons emitted from the
accretion disk (external Compton, EC). Multi-wavelength observational
studies from radio to $\gamma$-ray bands have indicated an interesting
feature in blazar SEDs; the synchrotron and Compton peak photon
energies decrease as the bolometric luminosity increases ({Fossati}
{et~al.}  1997, 1998; Ghisellini et al. 1998, {Donato} {et~al.}  2001,
{Ghisellini} \& {Tavecchio} 2008b, {Maraschi} {et~al.} 2008). This is
often called as the blazar SED sequence.  Although the validity of the
blazar sequence is currently still a matter of debate (e.g.,
{Padovani} {et~al.}  2007; {Ghisellini} \& {Tavecchio} 2008b;
{Maraschi} {et~al.} 2008), one can make a non-trivial prediction of
the EGRB spectrum if this blazar sequence is assumed. In other words,
the blazar SED sequence can be tested by comparing with the observed
EGRB spectrum.

In this paper we calculate the EGRB flux and spectrum from blazars, by
constructing a blazar GLF model that is consistent with the flux and
redshift distributions of the EGRET blazars, based on the LDDE scheme
and the blazar SED sequence. By introducing the blazar SED sequence,
we can make a reasonable and non-trivial prediction of the EGRB
spectrum for the first time, which can be compared with the observed
EGRB spectrum.

Recently, {Inoue}, {Totani}, \& {Ueda} (2008) has showed that EGRB in
the MeV band can naturally be explained by normal (i.e., non-blazar)
AGNs that compose the cosmic X-ray background.  This MeV background
component extends to $\sim 100$ MeV with a photon index of about 2.8,
by the Comptonization photons produced by nonthermal electrons in hot
coronae. Therefore, it should also contribute to the EGRB at
$\lesssim$ 1 GeV. We will investigate how much fraction of the
observed EGRB can be explained by the sum of the two components, i.e.,
non-blazar AGNs (dominant at $\lesssim$100 MeV) and blazars (dominant
at $\gtrsim$100 MeV).  

We will then make quantitative predictions for the {\it Fermi}
Gamma-ray Space Telescope (Atwood et al. 2009; formerly known as
GLAST) that has successfully been launched on June 11th, 2008, so that
our model can be tested quantitatively by the observational data
coming in the very near future \footnote{After we submit the first
  version of this paper (arXiv:0810.3580v1), the first {\it Fermi}
  catalog for bright gamma-ray sources including AGNs has appear (Abdo
  {et~al.}  2009a, b).  The sample size of blazars
  is about 100, which is bigger than the EGRET catalog by a modest
  factor of about 2.  We confine ourselves using only the EGRET data
  in this work, as the theoretical prediction in the pre-{\it Fermi}
  era.  A much larger number of blazars should be detected in the
  future {\it Fermi} sample, which should be compared with our
  predictions.}.

This paper is organized as follows. In \S \ref{sequence}, we will
present the treatment of the blazar SED sequence in our calculation.
We will describe the formulations for the blazar GLF model taking into
account the SED sequence in \S \ref{glf_model}, and the GLF parameter
determination by fitting to the EGRET data in \S \ref{glf_egret}. The
EGRB will be calculated and compared with the observed data in \S
\ref{section:egrb}, and we will present predictions for {\it Fermi} in
\S \ref{section:Fermi}.  Discussions on a few issues will be given in
\S \ref{section:discussion}, including a comparison of our results
with those in previous studies. Conclusions will
then be given in \S \ref{section:conclusions}. Throughout this paper,
we adopt the standard cosmological parameters of
($h,\Omega_M,\Omega_\Lambda$)=(0.7,0.3,0.7).

\section{The Blazar Population and the SED Sequence}
\label{sequence}

\begin{figure*}
  \begin{center}
\epsscale{}
\centering
\plotone{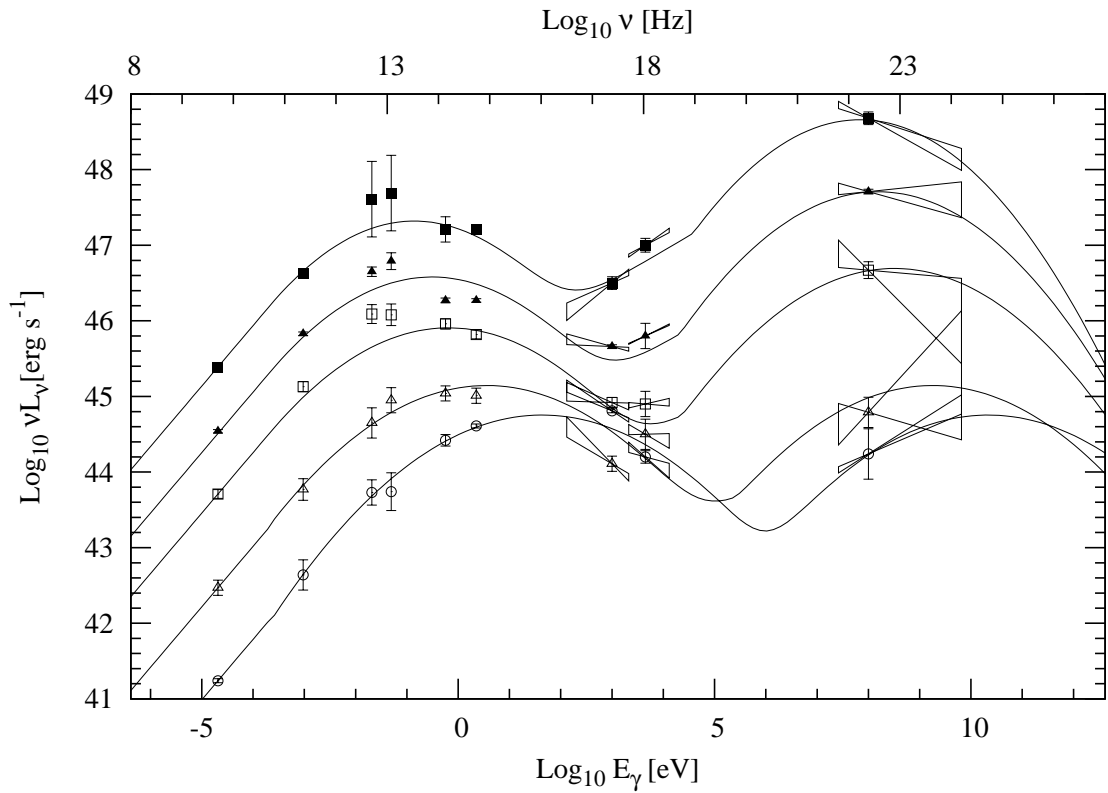}
\caption{The blazar SED sequence. The data points are the average SED
  of the blazars studied by {Fossati} {et~al.}  (1998) and D01. The
  solid curves are the empirical SED sequence models constructed and
  used in this paper. The model curves corresponds to the bolometric
  luminosities of $\log_{10} (P/{\rm erg \ s^{-1}}) = $ 49.50, 48.64,
  47.67, 46.37, and 45.99 (from top to bottom).}
\label{fig.sed}
\end{center}
\end{figure*}

Blazars are defined as the combined class of the two populations of
AGNs: BL Lac objects and flat spectrum radio quasars (FSRQs) (Urry \&
Padovani 1995)\footnote{Sometimes FSRQs are also called as optically
  violent variable (OVV) quasars or highly polarized quasars (HPQs),
  but they are essentially the same populations.}.  BL Lacs are
defined as AGNs whose nuclear non-thermal continuum emission is so
strong that the rest-frame equivalent width (EW) of the strongest
optical emission line is narrower than 5 \AA.  Such domination of
continuum is interpreted as the jet being directed towards the
observer; the continuum from jet is extremely enhanced by the
relativistic beaming compared with more isotropic line emission from
the regions around accretion disks.  FSRQs are AGNs with emission line
EW greater than 5 \AA, but whose spectral index in radio band is less
than $\alpha_r < $ 0.5, where $\alpha_r$ is defined by $f_\nu \propto
\nu^{-\alpha_r}$ (see Urry \& Padovani 1995 for detailed reviews).
Although FSRQs show discernible emission lines, the flat radio
spectrum is also an evidence of the dominating jet component. In fact
FSRQs show similar properties to BL Lacs, such as rapid time
variability and strong polarization.  These properties indicate that
BL Lacs and FSRQs are physically similar populations, and hence they
are often combined into a single class of blazars. The parameter that
discriminates BL Lacs and FSRQs is likely to be the luminosity;
generally FSRQs have larger luminosities than BL Lacs ({Fossati}
{et~al.} 1998, Ghisellini et al. 1998) \footnote{A recently popular
  classification is that blazars are divided into quasar-hosted
  blazars (QHB) that are the same population as FSRQ, low frequency BL
  Lacs (LBLs), and high frequency BL Lacs (HBLs), in the decreasing
  order of the absolute luminosity (Kubo et al. 1998).}.  In this
paper we also treat the BL Lacs and FSRQs as the single population of
blazars \footnote{After the submission of the first version of our
  paper (arXiv:0810.3580v1), Ghisellini et al. (2009) has shown that
  this treatment is adequate by using the latest data of {\it
    Fermi}. }.

{Fossati} {et~al.} (1997, 1998) and {Donato} {et~al.} (2001, hereafter
D01) constructed an empirical blazar SED model to describe the SED
sequence, based on fittings to observed SEDs from radio to
$\gamma$-ray bands. These models are comprised of the two components
(synchrotron and IC), and each of the two is described by a linear
curve at low photon energies and a parabolic curve at high energies,
in the plane of $\log_{10} (\nu L_\nu)$ and $\log_{10} \nu$.

Here we construct our own SED sequence model mainly based on the SED
model of D01, because there is a mathematical discontinuity in the
original D01 model. In the D01 model, two different mathematical
fitting formulae are used below and above the luminosity $\nu
L_\nu=10^{43}$ erg s$^{-1}$ at 5 GHz, and the luminosity of the
inverse-Compton component suddenly changes with a jump at this
critical luminosity. Our own SED sequence formula is described in
Appendix in detail, and the discontinuity is avoided there.  Once the
blazar luminosity is specified at a reference frequency (e.g., 5 GHz
in radio band), this empirical model gives the full blazar SED from
radio to gamma-ray bands.  In Fig.\ref{fig.sed} we show this empirical
blazar SED sequence model in comparison with the observed SED data
({Fossati} {et~al.} 1998; D01). It should be noted that the observed
data are means of many blazars in a certain luminosity range, and
there may be scatter of individual blazar SEDs from the sequence.
When the bolometric blazar luminosity $P \equiv \int \nu L_\nu$ is
specified, the blazar luminosity per unit frequency, $L_\nu(\nu, P)$,
is determined for any photon frequency $\nu$ by the blazar sequence
model. Note that, although blazars must be strongly anisotropic
emitters, $L_\nu$ is defined as the isotropic-equivalent luminosity
and hence the observed energy flux per unit frequency is given by
$F_\nu[\nu/(1+z)] = (1+z) \, L_\nu(\nu, P) / [4 \pi d_L(z)^2]$, where
$z$ is redshift and $d_L$ is the standard luminosity distance.

\section{The Model of Gamma-ray Luminosity Function of Blazars}
\label{glf_model}

\subsection{Relating Jet Power and X-ray Luminosity of AGNs}
\label{section:jet_power}

The cosmological evolution of X-ray luminosity function of AGNs has
been investigated intensively [{Ueda} {et~al.} 2003 (hereafter U03);
{Hasinger}, {Miyaji}, \& {Schmidt} 2005 (hereafter H05); {Sazonov}
{et~al.}  2007; {Gilli}, {Comastri}, \& {Hasinger} 2007]. These
studies revealed that AGN XLF is well described by the LDDE model, in
which peak redshift of density evolution increases with AGN
luminosity. Here we construct two models of blazar GLF based on the
two XLFs derived by U03 (in hard X-ray band) and H05 (in soft X-ray
band), both of which are based on the LDDE scheme.  The use of LDDE in
blazar GLF has been supported from the EGRET blazar data, because NT06
found that the EGRET data agrees with the LDDE model better than PLE
models.  However, the validity of this assumption from a theoretical
viewpoint should also be examined.

It is known that radio jet emission is linked to the dissipation
process in the accretion disk (Falcke \& Biermann 1995; Falcke et
al. 1995; Merloni et al. 2003; Falcke et al. 2004; K{\"o}rding et
al.2006). Therefore, it is natural to expect that power of blazar jet
is correlated with mass accretion rate onto super massive black holes,
which is also correlated with the X-ray luminosity from accretion
disk. Therefore we simply assume that the bolometric luminosity of
radiation from jet, $P$, is proportional to disk X-ray luminosity,
$L_X$.  It should be noted that $L_X$ is {\it not} the observed X-ray
luminosity of a blazar having a jet luminosity $P$; when an AGN is
observed as a blazar (i.e., the jet directed to an observer), its
X-ray flux is dominated by the radiation from the jet that is much
brighter than that from the accretion disk.  Rather, $L_X$ is the
luminosity that would be observed from a direction different from the
jet.  

A constant $P/L_X$ ratio is realized when, e.g., the jet kinetic
luminosity ($P_k$) is efficiently dissipated into blazar bolometric
luminosity ($P$), and both $P_k$ and $L_X$ are proportional to the
mass accretion rate ($\dot m$).  One should, however, be careful about
the latter condition. Recent observations of X-ray binaries indicate
that $P_k$ is generally proportional to $\dot m$, but $L_X$ is not,
when the accretion rate is much lower than the Eddington limit (i.e.,
low Eddington ratio) ({Gallo} {et~al.} 2003; {Gallo} {et~al.}
2005). Such accretion disks are described by radiatively inefficient
accretion flows (RIAF) rather than the standard accretion disk, and
$L_X$ is roughly proportional to $\dot m^2$ in the RIAF regime ({Kato}
{et~al.} 1998; {Narayan} \& {Quataert} 2005).  The RIAF picture is
well consistent also with the accretion flow onto the supermassive
black hole of the Galaxy (i.e., Sgr A$^*$) (see, e.g., Totani 2006 and
references therein).

Accretion rates of X-ray bright AGNs used to derive the AGN XLF are
generally close to the Eddington limit, otherwise they are hardly
detected by X-ray observations because of the rapid decrease of X-ray
luminosity with decreasing Eddington ratio.  Therefore, our blazar GLF
should be considered as that for high-Eddington-ratio AGNs, and
low-Eddington-ratio AGNs in the RIAF regime could be missed in our
analysis. Such a low-Eddington-ratio population might have a
significant contribution to blazar GLF, because we expect $P_k \propto
\dot m$ in contrast to $L_X \propto \dot{m}^2$ in the RIAF mode.
However, the black hole mass function predicted by time integration of
X-ray AGN LF is consistent with the local black hole mass function
inferred from the black-hole-mass versus bulge-mass relation,
indicating that black hole mass grows mainly in the high Eddington
ratio phase (e.g., {Marconi} {et~al.}  2004). If this is correct,
cosmic background radiation from jet activities should also be
dominated by AGNs in high Eddington ratio phase. Hence, it is
reasonable to expect that a significant part of EGRB flux can be
accounted for by blazars with high Eddington ratio phase, whose GLF
evolution is described by LDDE.

\subsection{Model Formulations}
\label{model_formulations}

In this paper we describe the blazar GLF in terms of $\nu L_\nu$
luminosity at a reference rest-frame photon energy $\epsilon_{\rm ref,
  res} \equiv$ 100 MeV, i.e., $L_\gamma \equiv (\epsilon_{\rm ref,
  res} /h_p) \ L_\nu(\epsilon_{\rm ref, res}/h_p, P)$, where $h_p$ is
the Planck constant.  According to the assumption justified in \S
\ref{section:jet_power}, we simply relate the bolometric blazar
luminosity $P$ and disk X-ray luminosity by the parameter $q$, as:
\begin{equation}
P=10^qL_X \ .
\label{power}
\end{equation}
Here, we define the disk luminosity $L_X$ to be that in the rest-frame
2--10 and 0.5--2 keV bands for the hard XLF (U03) and the soft XLF
(H05), respectively. Thus, $L_\gamma$ and $L_X$ have been related
through $P$.

\begin{deluxetable}{ccrrrrrrrrcrl}[b]
\tabletypesize{\scriptsize}[h]
\tablecaption{The parameters of the AGN XLF \label{XLF-parameters}}
\tablewidth{0pt}
\tablehead{
\colhead{} & \colhead{Ueda et al. 2003} & \colhead{Hasinger et al. 2005}
}
\startdata
$A_X$$^a$ & $5.04\times10^{-6}$ & $2.62\times10^{-7}$ \\
$\log_{10}$$L^*_X$$^b$ & $43.94_{-0.26}^{+0.21}$ & $43.94\pm 0.11$\\
$\gamma_1$ & $0.86\pm 0.15$ & $0.87\pm0.10$\\
$\gamma_2$ & $2.23\pm 0.13$ & $2.57\pm 0.16$\\
$z_c^*$ & $1.9^c$ & $1.96\pm 0.15$\\
$\log_{10}$$L_a$$^b$ & $44.6^c$ &  $44.67^c$\\
$\alpha$ & $0.335\pm 0.07$ &  $0.21\pm 0.04$\\
$p_1^*$ & $4.23\pm0.39$ & $4.7\pm0.3$\\
$p_2^*$ & $-1.5^c $ & $-1.5\pm0.7$\\
$\beta_1$ & $0.0^d$ & $0.7\pm 0.3$\\
$\beta_2$ & $0.0^d$ & $0.6\pm0.8$
\enddata
\tablenotetext{a}{In units of Mpc$^{-3}$.}  
\tablenotetext{b}{In units of ergs s$^{-1}$.}
\tablenotetext{c}{These quantities are treated as
fixed parameters in each XLF model.
}
\tablenotetext{d}{The indices $\beta_1$, $\beta_2$ are treated as 
constants in U03.
\vspace{0.5cm}}
\end{deluxetable}


The blazar GLF $\rho_\gamma$
is then obtained from the AGN XLF $\rho_X$, as
\begin{equation}
\rho_\gamma(L_\gamma, z) = \kappa\frac{dL_X}{dL_\gamma}\rho_X(L_X, z),
\end{equation}
where $\rho_\gamma$ and $\rho_X$ are the comoving number densities per
unit gamma-ray and X-ray luminosity, respectively.  The parameter
$\kappa$ is a normalization factor, representing the fraction of AGNs
observed as blazars.  In our GLF model, we adopt the same form in U03
and H05 for $\rho_X$, as:
\begin{equation}
\rho_X(L_X, z) = \rho_X(L_X,0)f(L_X,z),
\end{equation}
where $\rho_X(L_X, 0)$ is the AGN XLF at present.  This is
characterized by the faint-end slope index $\gamma_1 $, the bright-end
slope index $\gamma_2$, and the break luminosity $L_X^*$, as:
\begin{equation}
\rho_X(L_X,0)=\frac{A_X}{L_X \ {\rm ln}(10)} \left[ \left( \frac{L_X}{L_X^*} 
\right)^{\gamma_1} + \left( \frac{L_X}{L_X^*} \right)^{\gamma_2} \right]^{-1} \ ,
\end{equation}
where $A_X$ is the normalization 
parameter\footnote{The factor of $(L_X \ln 10)^{-1}$ 
appears because $A_X$ is defined as the pre-factor of the 
number density per unit $\log_{10} L_X$.} having a dimension of
volume$^{-1}$.
The function $f(L_X,z)$ describes the density
evolution, which is given by the following forms:
\begin{eqnarray}  
  f(L_X,z)=\left\{\begin{array}{ll}
      (1+z)^{p_1} & \text{$z \le z_c(L_X)$,} \\
      f \left( L_X,z_c(L_X) \right) 
      \left( \frac{1+z}{1+z_c(L_X)} \right)^{p_2} & \text{$z > z_c(L_X)$,} \\
    \end{array}\right.
\end{eqnarray} 
where $z_c$ is the redshift of evolutionary peak, given as
\begin{eqnarray}
z_c(L_X)=\left\{\begin{array}{ll}
    z_c^* & \text{$L_X \ge L_a$,} \\
    z_c^*(L_X/L_a)^\alpha & \text{$L_X < L_a$,} \\
    \end{array}\right.
\end{eqnarray} 
The evolutionary indices $p_1$ and $p_2$ are described by
using the parameters $p_1^*$, $p_2^*$, $\beta_1$, and $\beta_2$:
\begin{eqnarray}
p_1 = p_1^* + \beta_1 ({\rm \log_{10}} L_X -44.0), \\
p_2 = p_2^* + \beta_2 ({\rm \log_{10}} L_X -44.0).
\end{eqnarray} 
The parameters obtained by the fit to the observed data of X-ray AGNs
in U03 and H05 are shown in Table.\ref{XLF-parameters}.

When we calculate the EGRB flux, it diverges if $\gamma_1 > 1$ and GLF
is integrated down to $L_\gamma \rightarrow 0$.  Therefore we set the
minimum of the gamma-ray luminosity as $L_{\gamma,\rm min}$=10$^{43}$
erg s$^{-1}$ in the EGRB calculation, because we will find that there
is no EGRET blazars below this value (see Fig.  \ref{EGRET_dist}).
However, it should be kept in mind that there might be a considerable
contribution to the EGRB flux from blazars below $L_{\gamma, \min}$
when $\gamma_1 > 1$.

\section{Gamma-Ray Luminosity Function Determined 
by the EGRET Blazar Data}
\label{glf_egret}

\subsection{The Maximum Likelihood Method}
\label{likelihood}

We use the maximum likelihood method to search for the best-fit model
parameters of the blazar GLF to the distributions of the observed
quantities of the EGRET blazars (gamma-ray flux and redshift).
The analysis method and the data used are essentially the same
as those  in NT06. 

Observed gamma-ray flux $F_\gamma$ of EGRET blazars are photon flux
at $\epsilon_\gamma \ge \epsilon_{\rm min, obs} 
\equiv 100 \ {\rm MeV}$ in photons cm$^{-2}$
s$^{-1}$, where $\epsilon_{\rm min, obs}$ is in the observer's frame. 
For a given redshift, $F_\gamma$ can be
related to $P$ by the blazar SED sequence model
as follows:
\begin{equation}
F_\gamma = \frac{1+z}{4\pi d_L(z)^2} 
\int_{ \epsilon_{\rm min, obs} (1+z)/h_p}^{\infty}
\frac{L_\nu(\nu, P)}{h_p \nu} d\nu \ .
\label{eq_Fg}
\end{equation}
Since $P$ has been related to the gamma-ray luminosity $L_\gamma$ by
the SED sequence, one can calculate $L_\gamma(z, F_\gamma)$ for a
given set of $z$ and $F_\gamma$ through $P$.

A specified GLF model predicts the distribution function
$d^3N/dzdF_\gamma d\Omega$ of redshift $z$, flux $F_\gamma$, and the
location of a blazar in the sky specified by a solid angle $\Omega$, 
which is given as:
\begin{eqnarray}
  \frac{d^3N(z,F_{\gamma},\Omega)}{dzdF_\gamma d\Omega}
  = \frac{dV}{dz}\rho_\gamma(L_\gamma,z)  \epsilon(F_\gamma, z) \\
  \times \Theta\lfloor F_\gamma-F_{\gamma,{\rm lim}}(\Omega)\rfloor,\nonumber
\end{eqnarray} 
where $dV/dz$ is the comoving volume element per unit
solid angle, $\Theta$ the step function ($\Theta(x) = 1$ and 0 for $x
\geq$ 0 and $x < 0$, respectively), and $F_{\gamma,{\rm lim}}(\Omega)$
the sensitivity limit of EGRET for point sources that is a function of
the Galactic latitude.  The functional form of $F_{\gamma,
  \lim}(\Omega)$ is given in NT06.  The detection efficiency
$\epsilon(F_\gamma, z)$ represents the identification probability as a
blazar by finding a radio counterpart. This is described \S 2.3 in
NT06, but here we modified the relation between the radio luminosity
$L_R$ and $L_\gamma$ from a simple linear relation in NT06 to that
predicted from our blazar SED sequence model. As in NT06, we also take
into account a log-normal scatter around the mean relation of
$L_R/L_\gamma$ with a standard deviation of $\sigma_p = 0.49$ in
$\log_{10} (L_R/L_\gamma)$.  It should be noted that the blazar SED
sequence is an averaged SED for groups of blazars binned by radio
luminosity, and some scatter of $L_R/L_\gamma$ is expected for
individual blazars.

The likelihood function is given by 
\begin{equation}
  \mathcal{L}=\prod_{i=1}^{N_{{\rm obs}}}\left\lfloor 
\frac{1}{N_{{\rm exp}}}\frac{dN^3(z_i,F_{\gamma,i},\Omega_i)}{
dz \, dF_\gamma \, d\Omega}
 \right\rfloor \ ,
\end{equation}
where the subscript $i$ is identification number of each blazar, 
$N_{\rm obs}$ the observed number of blazars,
and $N_{{\rm
    exp}}$ the expected number of the blazar detections, i.e.,
\begin{equation}
N_{{\rm exp}}=\int dz \int dF_\gamma \int d\Omega \,
\frac{d^3N}{dz\, dF_\gamma \, d\Omega}.
\end{equation}
The likelihood function does not depend on the normalization of GLF,
and the normalization parameter $\kappa$ is determined by requiring
$N_{\exp} = N_{\rm obs}$. There are $N_{\rm obs} =46$ blazars in the
sample analyzed in NT06 and this work.

\begin{figure*}
  \begin{center}
\includegraphics[width=180mm]{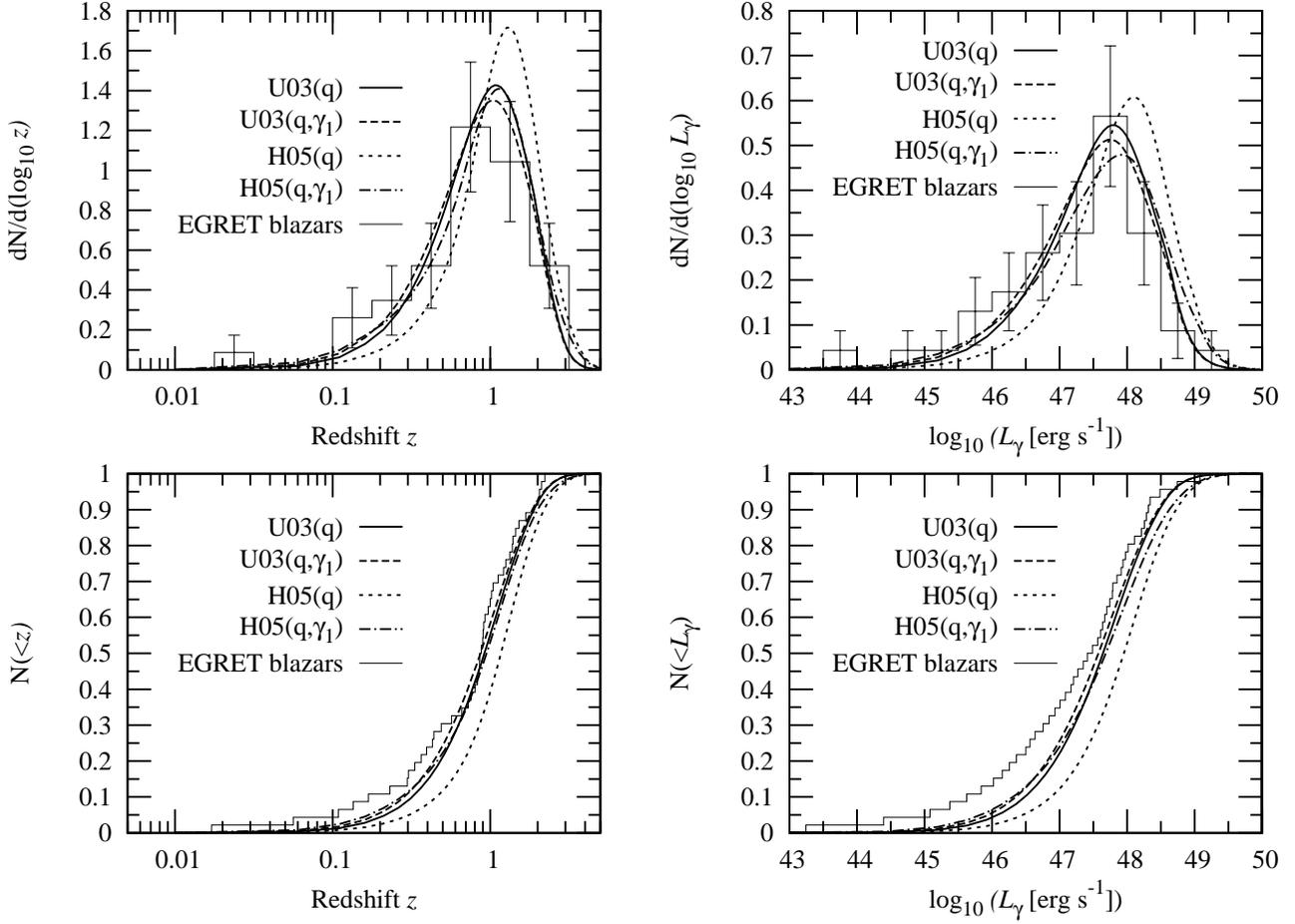}
    \caption{Top panels: redshift and gamma-ray luminosity ($\nu
      L_\nu$ at 100 MeV) distributions of EGRET blazars. The histogram
      is the binned EGRET data, with one sigma Poisson errors.  The
      four model curves are the best-fits for the GLF models of
      U03($q$), H05($q$), U($q,\gamma_1$), and H05($q,\gamma_1$), as
      indicated in the figure.  Bottom panels: the same as the top
      panels, but for cumulative distributions.}
    \label{EGRET_dist}
  \end{center}
\end{figure*}

\subsection{The Best Fit Parameters}
\label{best_fit}

\begin{deluxetable}{ccrrrrrrrrcrl}[h]
\tabletypesize{\scriptsize}[t]
\tablecaption{Best-fit parameters for blazar GLF \label{best-fit-glf}}
\tablewidth{0pt}
\tablehead{
\colhead{} & \colhead{U03($q$)} & \colhead{U03($q$,$\gamma_1$)} 
  & \colhead{H05($q$)} & \colhead{H05($q$,$\gamma_1$)} }
\startdata
$q$ & 4.92$_{-0.10}^{+0.21}$ & $4.93_{-0.10}^{+0.25}$ & 5.29$_{-0.22}^{+0.26}$ 
  & $5.35_{-0.21}^{+0.25}$	\\
$\gamma_1$ & 0.86$^a$ & $0.93\pm 0.13$ & 0.87$^a$ & $1.11_{-0.12}^{+0.11}$	\\
$\kappa$ & $1.7\times10^{-6}$ & $1.5\times 10^{-6}$ & $9.5\times 10^{-6}$ & $6.0\times 10^{-6}$	\\
\cutinhead{KS test probabilities}
$z$ & 53.8\% & 86.9\% & 0.15\% & 33.0\%\\
L$_{\gamma}$ & 25.6\% & 48.4\% & 0.08\% & 28.5\%
\enddata

\tablecomments{The best-fit values of the model parameters ($q$,
  $\gamma_1$, $\kappa$) obtained from the maximum likelihood
  analysis. The KS probabilities of the best-fit models are also shown
  for the redshift and gamma-ray luminosity distributions in the last
  two rows.  }

\tablenotetext{a}{These parameters are fixed at the original AGN XLF
  values in these analysis, and the original values are
  shown.\vspace{0.5cm}}

\end{deluxetable}

In the first analyses, we take $q$ as the only one free parameter of
the blazar GLF, with all the XLF model parameters fixed at the values
of U03 and H05. These are hereafter called as U03($q$) and H05($q$)
fits, respectively. In the second analyses, we take the faint-end
slope index of XLF $\gamma_1$ as another free parameter in addition to
$q$. These are hereafter called as U03($q$, $\gamma_1$) and H05($q$,
$\gamma_1$) fits, respectively.  The motivation of these models is to
investigate the effect of the uncertainty about the faint-end slope,
because it may have significant effect on EGRB if $\gamma_1 \gtrsim
1$.  The best-fit values for these four fits are shown in
Table. \ref{best-fit-glf}.  Figure \ref{EGRET_dist} shows the
distributions of redshift and gamma-ray luminosity predicted by the
best-fit models, in comparison with the EGRET data.  We obtained the
best-fit values of $q \sim 5$, meaning that the observed jet
luminosity $P$ (bolometric blazar luminosity) is typically $10^5$
higher than the disk X-ray luminosity.  In the U03($q, \ \gamma_1$)
model, the characteristic AGN X-ray luminosity $L_X^*$ (the break in
XLF) corresponds to the blazar gamma-ray luminosity of $L_\gamma^* =
10^{48} \ \rm erg \ s^{-1}$.

\begin{figure*}
  \begin{center}
    \begin{tabular}{cc}
      \resizebox{90mm}{!}{\includegraphics{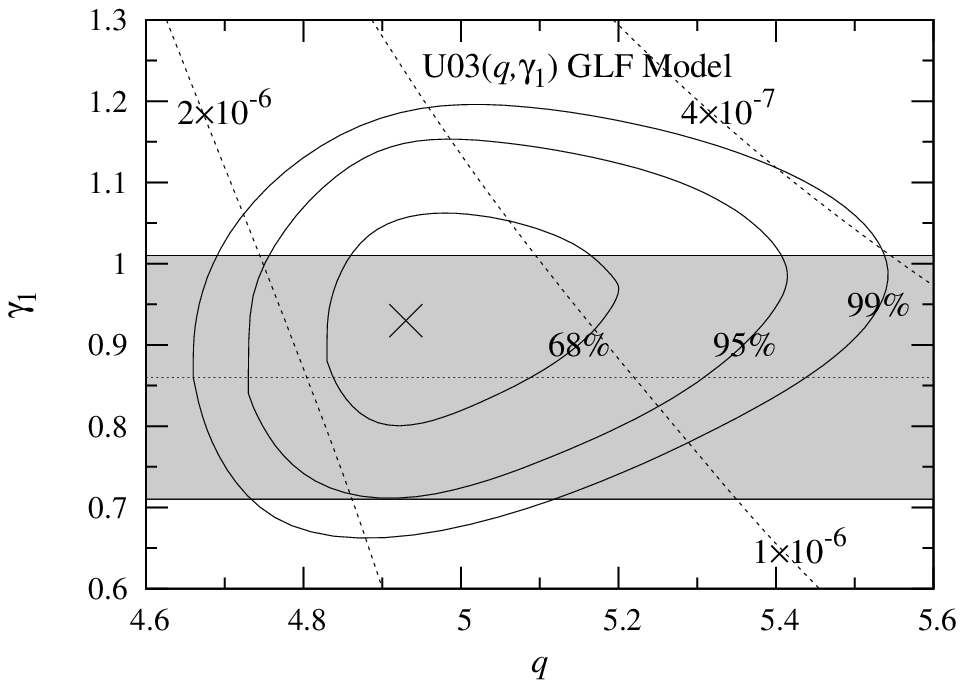}} &
      \resizebox{90mm}{!}{\includegraphics{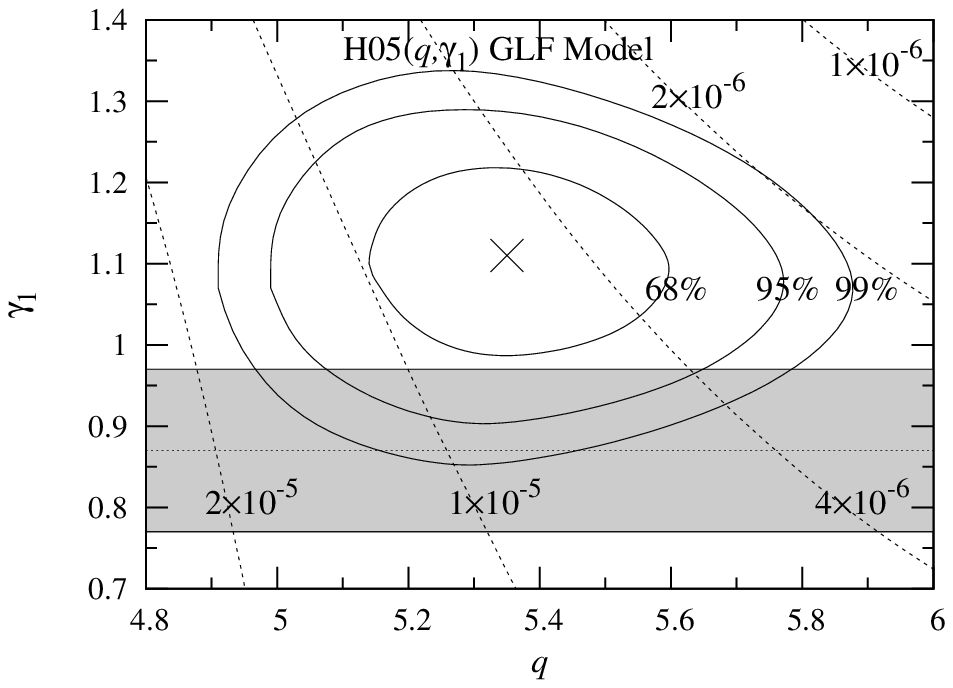}} \\
    \end{tabular}
    \caption{$Left$: Solid contours show the 68\%, 95\%, and $99\%$ CL
      likelihood contours for the LDDE model parameters [the faint-end
      slope index, $\gamma_1$, and the ratio of the blazar emission
      power to disk X-ray luminosity, $q$] in the case of the U03 XLF. The
      best-fit values ($q$,$\gamma_1$)=(4.93, 0.93) are shown by the
      cross.  The dotted contours are for the parameter $\kappa$, the
      normalization ratio of blazar GLF to AGN XLF. The $\kappa$
      values for the contours are indicated in the figure.  The shaded
      region indicates the $1 \sigma$ error region of the original $\gamma_1$
      value determined for AGN XLF by X-ray surveys.  $Right$:
      The same as the left panel, but for the H05 XLF. The best-fit
      values ($q$,$\gamma_1$)=(5.35, 1.11) are shown by the cross. 
      }
    \label{contour}
  \end{center}
\end{figure*}

We performed the Kolmogorov-Smirnov (KS) test to see the goodness of
fits for the best-fit results of each of the four fits, and the chance
probabilities of getting the observed KS deviation are shown in Table
\ref{best-fit-glf}.  Except for the H05$(q)$ model, the fits are
statistically acceptable.  Since the best KS test value is obtained
for the U03($q, \gamma_1$) GLF model, we use the U03($q$, $\gamma_1$)
GLF model as the baseline below, when only one of the four fits is
presented. In Fig. \ref{contour} we show the allowed regions of the
model parameters in the U03($q$, $\gamma_1$) and H05($q$, $\gamma_1$)
models. The best-fit value of $\gamma_1$ in the U03($q$, $\gamma_1$)
model is in good agreement with the original value derived by U03.
The value in the H05($q$, $\gamma_1$) model is slightly larger than
the original H05 value, but the statistical significance is not large.
The difference between the U03($q$, $\gamma_1$) and H05($q$,
$\gamma_1$) fits may be a result of different selections of AGNs in
soft (H05) and hard X-ray (U03) bands, because strongly obscured AGNs
would more easily be missed in soft X-ray bands than in hard X-ray
bands. 

It should be noted that the results of the acceptable KS probabilities
and the similar $\gamma_1$ values to the original XLFs give some
support to our basic assumption of a simple proportionality between
the X-ray accretion luminosity $L_X$ and the jet luminosity $P$. As
mentioned above, X-ray faint AGNs may be in the RIAF mode
(low-Eddington ratio) rather than the standard disk mode, and the
proportionality between $L_X$ and $P$ could be violated. Therefore our
results implies that the EGRET data can be described well with the
assumption that the majority of EGRET blazars are in the standard disk
mode or high Eddington ratio phase. {\it Fermi} will probe much
fainter blazars than EGRET, and we may see the deviation from the
proportionality between $L_X$ and $P$.

\section{The Gamma-Ray Background Spectrum}
\label{section:egrb}

\subsection{the EGRB Spectrum Calculation}
We calculate the EGRB spectrum by integrating our blazar SED sequence
model in the redshift and luminosity space, using the blazar GLF
derived in \S \ref{glf_model}. The spectrum of EGRB radiation (photon flux 
per unit photon energy and per steradian) is calculated as
\begin{eqnarray}
  \frac{d^2F(\epsilon_\gamma)}{d\epsilon_\gamma d\Omega}&=&\frac{c}{4\pi}
      \int_0^{z_{\rm max}}dz
        \frac{dt}{dz}
      \int_{L_{\gamma,\rm min}}^{L_{\gamma}(F_{\rm EGRET},z)}dL_\gamma \
  \rho_\gamma (L_\gamma, z) \\ \nonumber
   & \times & \frac{(1+z)}{h_p}
  \ \frac{L_\nu[\epsilon_\gamma (1+z) / h_p, P(L_\gamma)]}{
  \epsilon_\gamma (1+z) }\\ \nonumber
  &\times& e^{-\tau_{\gamma \gamma}}(z, \epsilon_\gamma) \ , 
\end{eqnarray}
where $\epsilon_\gamma$ is the observed gamma-ray photon energy, $t$
the cosmic time, and $dt/dz$ can be calculated by the Friedmann
equation in the standard cosmology.  We assume $z_{\max} = 5$ in this
calculation, but the EGRB flux is hardly dependent on this parameter,
since the peak of GLF/XLF evolution is well below $z \sim 5$.  The
minimum gamma-ray luminosity is set at $L_{\gamma, \min} = 10^{43} \
\rm erg \ s^{-1}$ as mentioned earlier. Since EGRET has already
resolved bright gamma-ray sources, we should not include those sources
in the EGRB calculation. Therefore, the maximum gamma-ray luminosity
in the integration should be $L_{\gamma}(F_{\rm EGRET},z)$, which is
the luminosity of the blazar having a flux $F_\gamma = F_{\rm EGRET}$
at a given $z$, where $F_{\rm EGRET}=7 \times 10^{-8} \ \rm photons \
cm^{-2} \ s^{-1}$ is the EGRET sensitivity above 100 MeV.

Very high energy photons ($\gtrsim$ 20 GeV) from high redshift are absorbed
by the interaction with the cosmic infrared background (CIB) radiation
({Salamon} \& {Stecker} 1998; {Totani} \& {Takeuchi} 2002; {Kneiske}
{et~al.} 2004; {Stecker}, {Malkan}, \& {Scully} 2006), and
$\tau_{\gamma \gamma}(z, \epsilon_\gamma)$ is the optical depth of the
universe against this reaction.  In this paper, we adopt the model of
{Totani} \& {Takeuchi} (2002) for CIB and $\tau_{\gamma \gamma}$.  The
gamma-ray absorption produces electron-positron pairs, and the pairs
would scatter the cosmic microwave background (CMB) radiation to make
the secondary contribution (the cascade component) to high energy
background radiation ({Aharonian} {et~al.} 1994; {Fan} {et~al.}
2004). We take into account this cascading emission to calculate EGRB
spectrum, considering only the first generation of created pairs, by
using the same formulations in {Kneiske} \& {Mannheim} (2008). In the
following results, we will find that the amount of energy flux
absorbed and reprocessed in intergalactic medium (IGM) is only a small
fraction of the total EGRB energy flux, and hence the model dependence
of CIB or the treatment of the cascading component does not have
serious effects on our conclusions in this work.

\subsection{The EGRB Spectrum from Blazars}
\label{composition}

\begin{figure*}
  \begin{center}
    \begin{tabular}{cc}
\includegraphics[width=180mm]{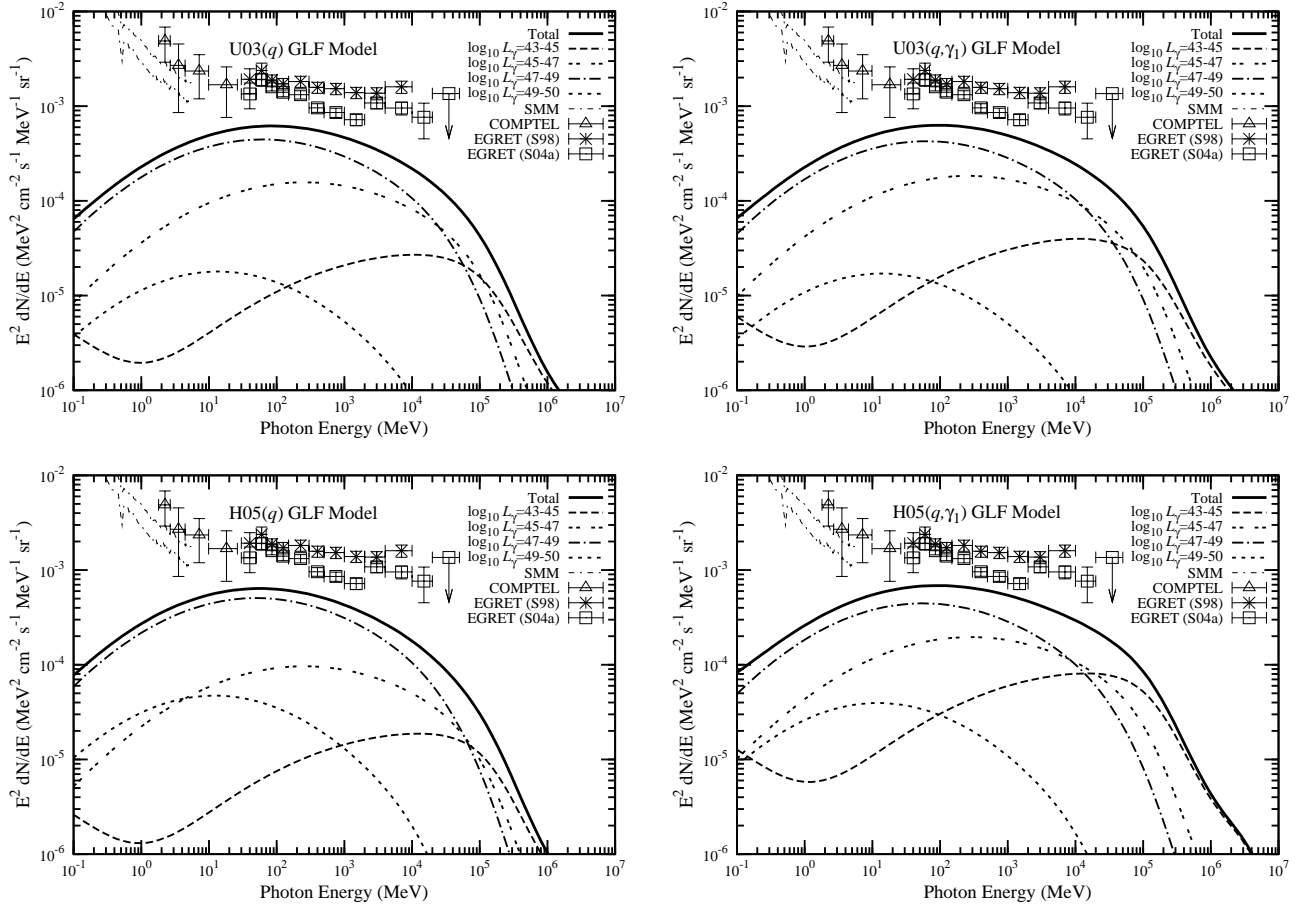}
    \end{tabular}
    \caption{The
      blazar EGRB spectrum (energy flux per unit logarithmic photon
      energy and per steradian).  The four panels are for the four
      different GLF models of U03($q$), U03($q,\gamma_1$), H05($q$),
      and H05($q,\gamma_1$). The solid curve is for the total EGRB
      spectrum from all blazars, and the other curves are for a
      particular range of blazar luminosity, as indicated in the
      figure ($L_\gamma$ in units of erg/s).  The effect of absorption
      by CIB is included, while the reprocessed cascade component is
      not included.  The observed data of SMM ({Watanabe} {et~al.}
      1999), COMPTEL ({Kappadath} {et~al.}  1996), and EGRET
      [{Sreekumar} {et~al.} 1998 (S98); {Strong} {et~al.} 2004a
      (S04a)] experiments are also shown with the symbols indicated in
      the figure.}
    \label{fig.blazar}
  \end{center}
\end{figure*}

Figure \ref{fig.blazar} shows the $\nu F_\nu$ EGRB spectrum predicted
by the best-fit GLF model parameters obtained above. Here, we show the
total EGRB flux as well as the contributions from different $L_\gamma$
ranges. This EGRB spectrum is the primary spectrum directly from
blazars, because in this calculation the effect of photon absorption
in IGM is included but the cascade emission is not.  The data of SMM
({Watanabe} {et~al.} 1999), COMPTEL ({Kappadath} {et~al.}  1996) and
EGRET ({Sreekumar} {et~al.}  1998; {Strong} {et~al.}  2004a) are also
shown. As seen in Figure \ref{fig.blazar}, the contribution from
low-luminosity blazars with $\log_{10} (L_\gamma/{\rm erg \ s^{-1}})$ =
43--45 is significant only above 1 GeV, because of the assumed SED
sequence of blazars.  The contribution from these low-luminosity
blazars is larger in the model of H05($q,\gamma_1$) than that of
H05($q$), because the faint-end slope of the GLF becomes steeper when
$\gamma_1$ is treated as a free parameter.

An important implication here is that the contribution from the
so-called MeV blazars, which have their SED peaks at around MeV, is
negligible in the MeV gamma-ray background flux, although such MeV
blazars have been suspected as a possible origin of the MeV
background. This is because MeV blazars have a large luminosity of
$\log_{10} (L_\gamma/{\rm erg \ s^{-1}})$ = 49--50 in the blazar SED
sequence, and the number density of such blazars is small.

\begin{figure*}
  \begin{center}
    \begin{tabular}{cc}
      \resizebox{90mm}{!}{\includegraphics{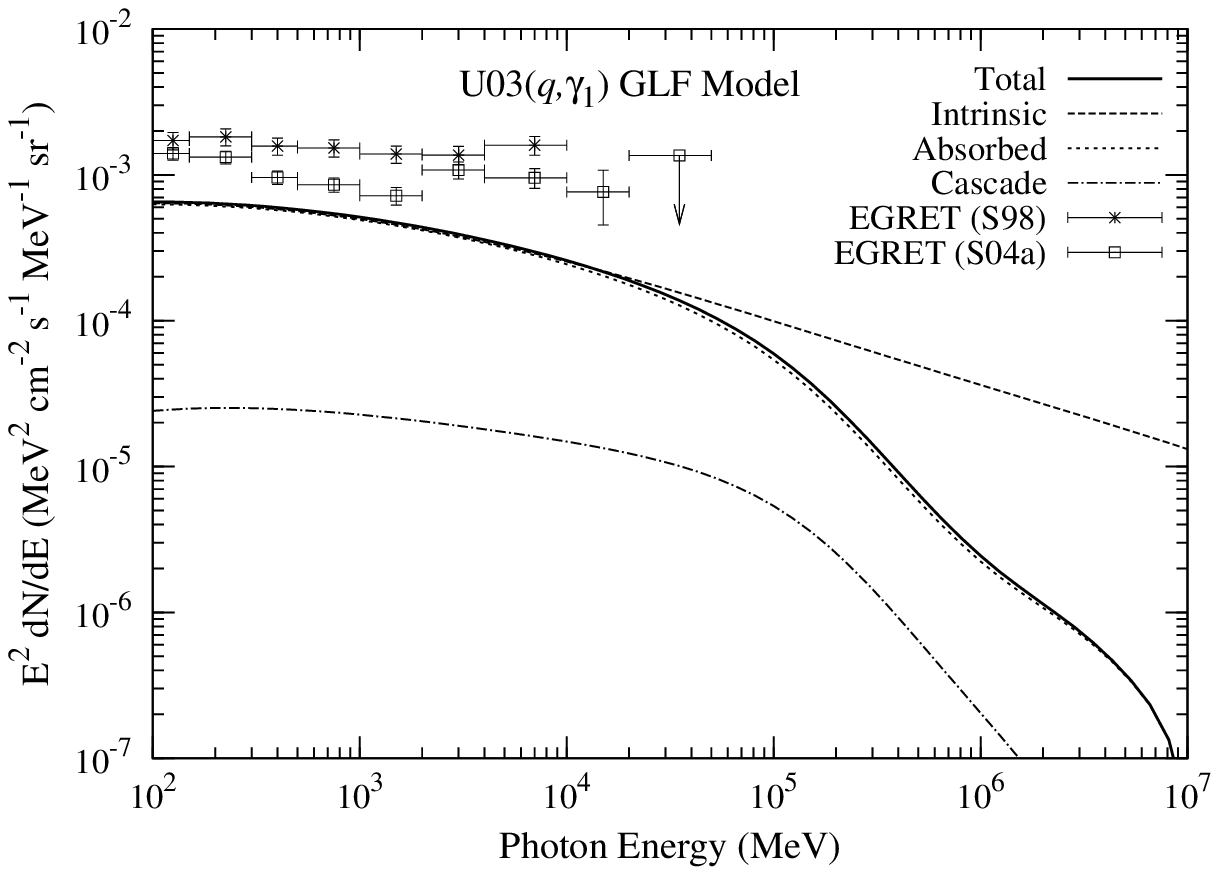}} &
      \resizebox{90mm}{!}{\includegraphics{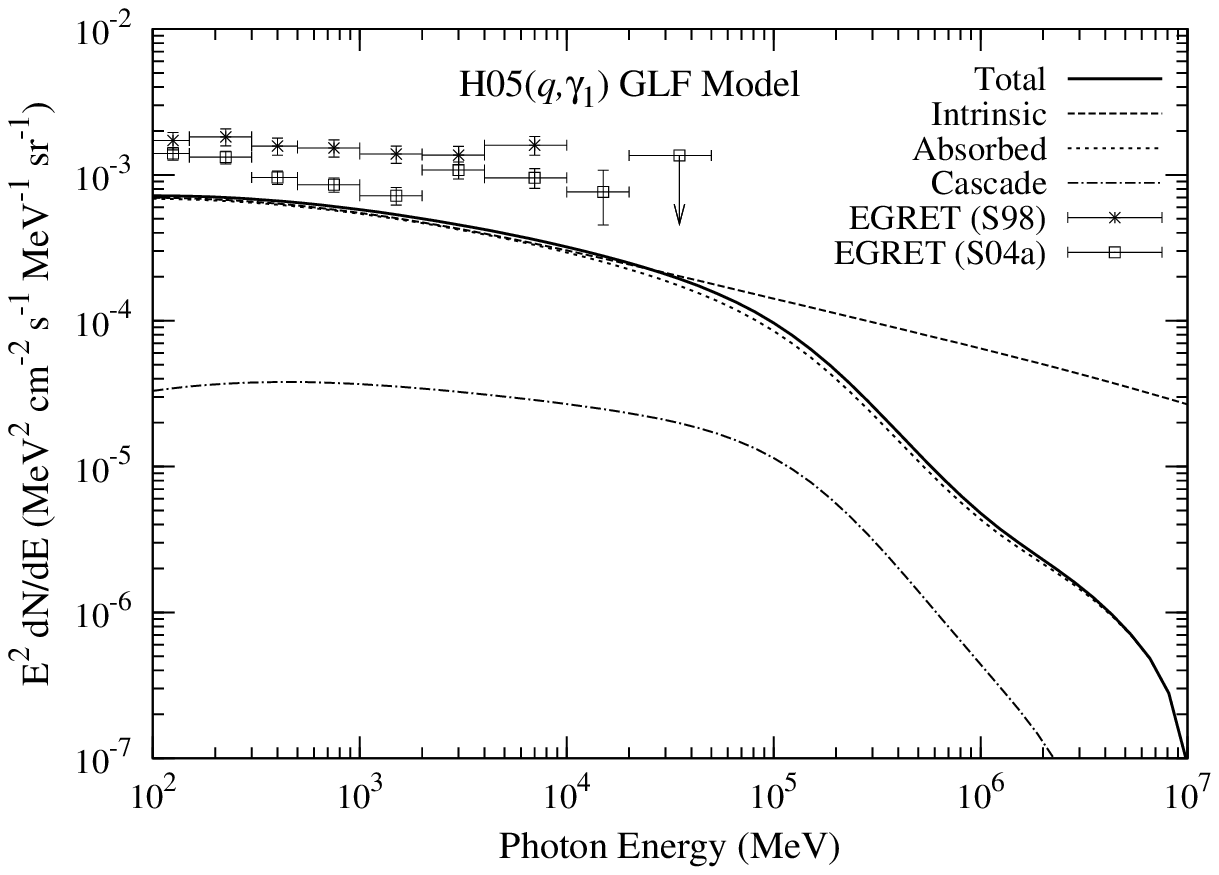}}
    \end{tabular}
    \caption{The blazar EGRB spectrum. The two panels are for the two
      different GLF models of U03($q,\gamma_1$) and
      H05($q,\gamma_1$). Three model predictions for the intrinsic (no
      absorption by CIB), absorbed, and cascade (reprocessed emission
      by electrons/positrons produced in IGM) components of the EGRB
      spectrum are shown by line markings shown in the figure.  The
      solid curve is the total flux, i.e., absorbed plus cascade
      components.  The EGRET data are the same as those in
      Fig. \ref{fig.blazar}. 
 }
    \label{fig.cascade}
  \end{center}
\end{figure*}

Figure \ref{fig.cascade} shows the intrinsic (the spectrum without
taking into account the absorption by CIB), absorbed (the same as
Fig. \ref{fig.blazar}), and cascading components of EGRB spectrum,
as well as the total (absorbed+cascade) spectrum.  The
absorption by CIB becomes significant at $\epsilon_\gamma \gtrsim 100$
GeV, and the absorbed EGRB photons are converted into lower energy
gamma-rays, with the energy flux roughly conserved.  However, the
absorbed energy flux of EGRB above $\sim$ 100 GeV is not significantly
larger than the unabsorbed energy flux, and hence the increase of EGRB
flux or change of the EGRB spectrum at $\lesssim$ 100 GeV by the
cascading component is not a large effect.

\subsection{The EGRB Spectrum from Non-blazar AGNs}
\label{I08}

In addition to blazars, we take into account non-blazar AGNs as the
source of EGRB.  Inoue et al. (2008, hereafter ITU08) has shown that
non-blazar AGNs are a promising source of EGRB at $\sim$1--10 MeV.  In
this scenario, a nonthermal power-law component extends from hard
X-ray to $\sim$10 MeV band in AGN spectra, because of nonthermal
electrons that is assumed to exist ubiquitously in hot coronae around
AGN accretion disks. The existence of such nonthermal electrons is
theoretically reasonable, because the hot coronae are the essential
component in the standard picture of X-ray emission from AGNs, and
magnetic reconnection is the promising candidate for the heating
source of the hot coronae ({Laor} \& {Behar} 2008).  Magnetic
reconnections should produce nonthermal particles, as is well known in
e.g., solar flares or the earth magnetosphere.  Although several
sources have been proposed as the origin of the MeV background, this
model gives the most natural explanation for the observed MeV
background spectrum that is a simple power-law smoothly connected to
CXB.

Theoretically, there is no particular reason to expect a cut-off of
the nonthermal emission around 10 MeV, and it is well possible that
this emission extends beyond 10 MeV with the same power index.  Then,
this component could make some contribution to EGRB in the EGRET
energy band.  The EGRB spectrum from the non-blazar AGNs calculated
based on the model of ITU08 is shown in Figure \ref{fig.egrb}, in
comparison with the blazar component.  The model parameters of ITU08
have been determined to explain the EGRB in the MeV band.  There is
some discrepancy between the MeV EGRB data of SMM and COMPTEL, and the
reason for this is not clear.  Here we use two model parameter sets of
$(\Gamma, \gamma_{\rm tr})=(3.5,4.4)$ and $(3.8,4.4)$ in ITU08, where
$\Gamma$ is the power-law index of nonthermal electron energy
distribution and $\gamma_{\rm tr}$ is the transition electron Lorentz
factor above which the nonthermal component becomes dominant.  These
two parameter sets are chosen so that they fit to the COMPTEL and SMM
data, respectively.  The EGRB flux is dominated by blazars at the
energy range above $\sim$100 MeV, while the non-blazar component is
dominant at the lower photon energies.

\subsection{On the Origin of EGRB}
\label{subsec:origin_egrb}

\begin{figure*}
  \begin{center}
    \begin{tabular}{cc}
      \resizebox{90mm}{!}{\includegraphics{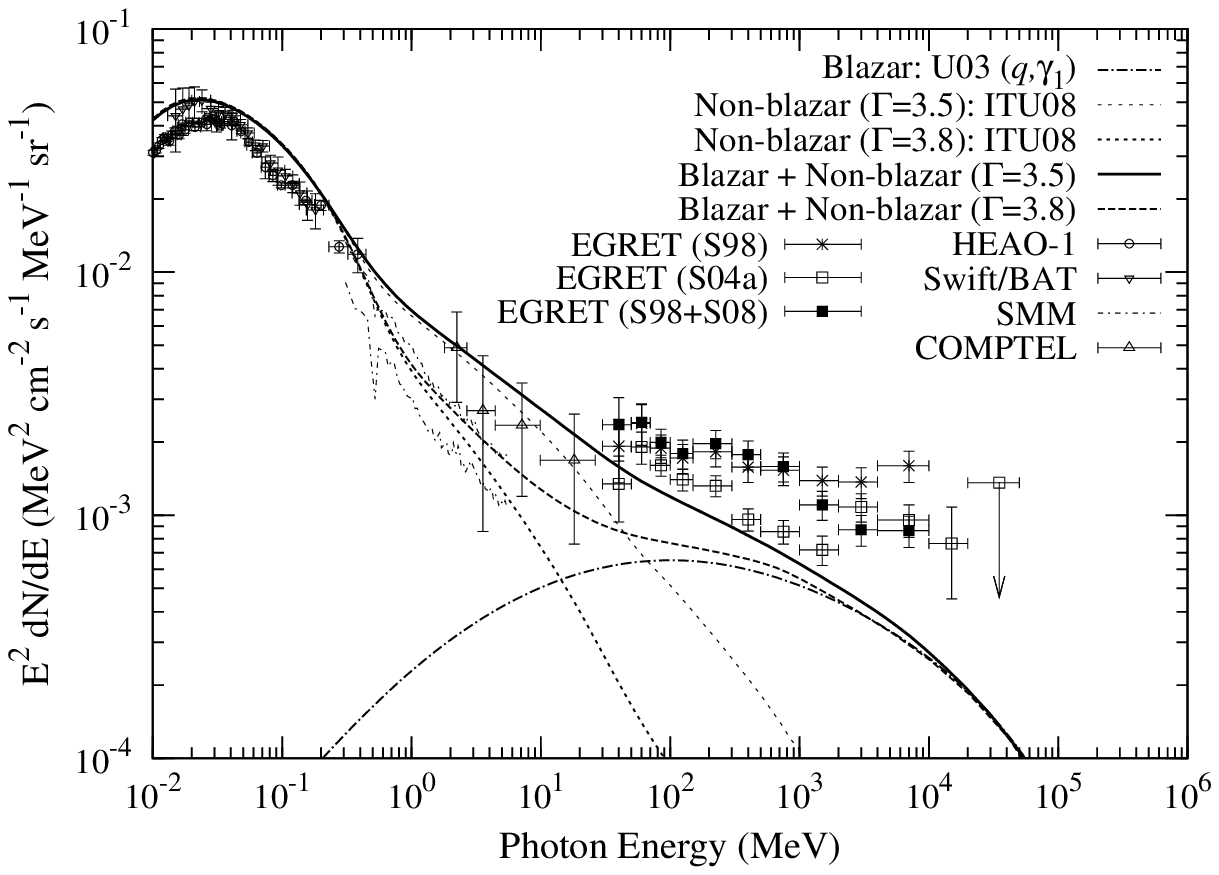}} &
      \resizebox{90mm}{!}{\includegraphics{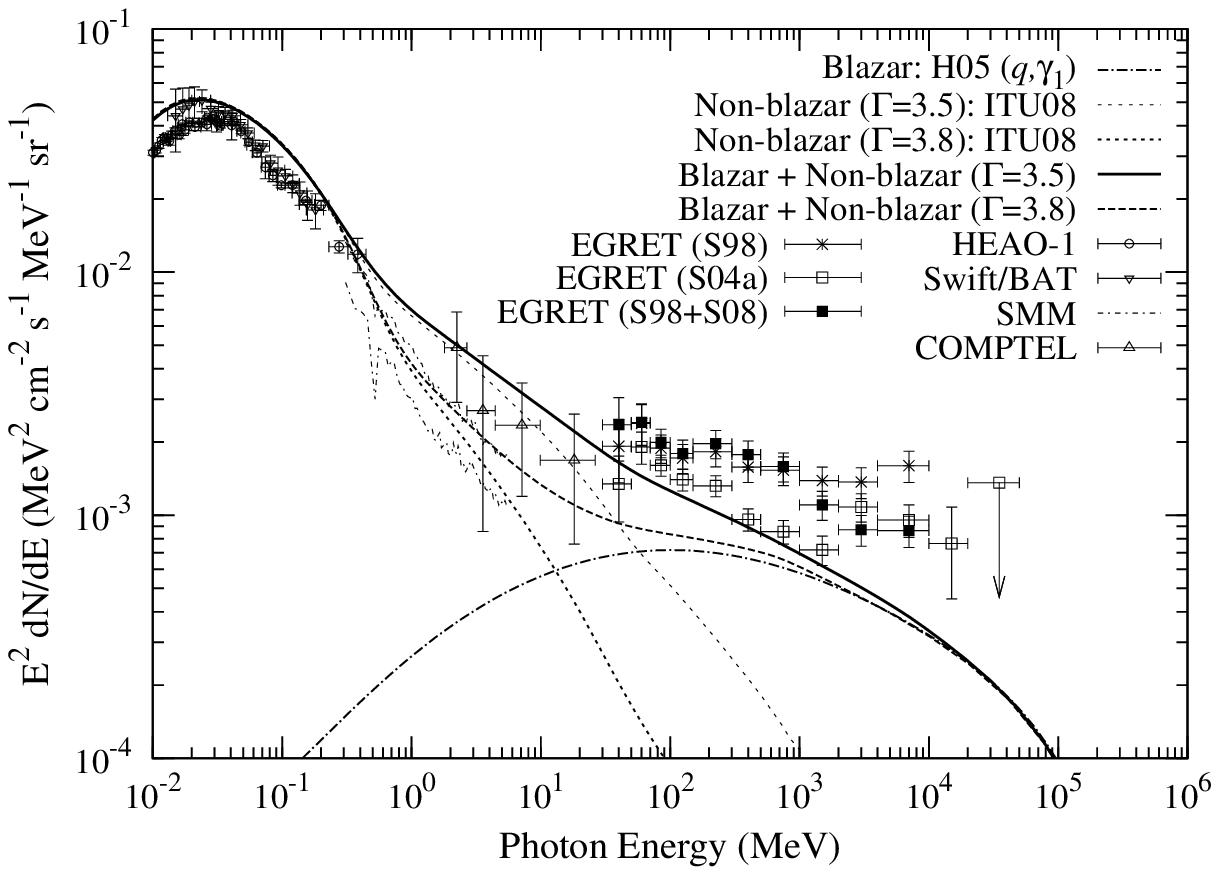}}
    \end{tabular}
    \caption{The EGRB spectrum from non-blazar AGNs and blazars.  The
      two panels are for the two different blazar GLF models of
      U03($q,\gamma_1$) (left) and H05($q,\gamma_1$) (right). The
      model curves of the blazar component (absorbed+cascade),
      non-blazar AGN component, and the total of the two populations
      are shown. Note that two models are plotted for the non-blazar
      component with different values of $\Gamma$ (see the
      line-markings indicated in the figure).  The observed data of
      HEAO-1 ({Gruber} {et~al.} 1999) and
 {\it Swift}/BAT ({Ajello} {et~al.} 2008) are shown. We also plot a new EGRET
      data denoted as ``S98+S08'', which is the original EGRET
      determination of {Sreekumar} {et~al.} (1998) corrected by the
      correction factors proposed by Stecker et al. (2008, S08).  The other
      data are the same as those in Fig. \ref{fig.blazar}.
      }
    \label{fig.egrb}
  \end{center}
\end{figure*}

Now we compare the EGRB data above 100 MeV with our model predictions.
Because of the uncertainties in the EGRB measurements, we plotted
three types of EGRB EGRET data in this figure (see below for the explanations of the data points).  As can
be seen in Fig. \ref{fig.egrb}, the overall background spectrum from
X-ray to 1 GeV predicted by our blazar plus non-blazar model is
similar to the observed data. Especially, the EGRB prediction using
the ITU08 non-blazar background model with $\Gamma = 3.5$ is in nice
agreement with the observed data of Strong et al. (2004a) in 0.1--1
GeV. In this case, the predicted EGRB flux at 100 MeV can account for
80\% of the observed flux, which is a considerably higher fraction
than those in previous studies (Chiang \& Mukherjee 1998; M\"ucke \&
Pohl 2000; NT06).  It should be noted that the contribution to the
EGRB flux at 100 MeV from blazars is $\sim$45 \%, which is similar to
those estimated by previous studies. The high fraction is due to the
addition of the EGRB component from non-blazar AGNs.

The prediction is still 20 \% short of the observed flux, and the
discrepancy becomes more serious when the model is compared with the
Strong et al. (2004a) data above 1 GeV or with the original EGRB
determination by the EGRET team (Sreekumar et al. 1998). Especially,
the apparent excess of the observed EGRB flux beyond 1 GeV might
be a contribution from a completely different component, e.g.,
dark matter annihilation (e.g., Oda, Totani, \& Nagashima 2005).
Therefore we carefully discuss the possible origin of the discrepancy
below.

We should first examine the uncertainties in the model
prediction. Since the GLF likelihood determination is based only on 46
blazars, there is a statistical uncertainty of 15 and 32\% in the
normalization of GLF and the EGRB flux at 68 and 95\% C.L.,
respectively. The sensitivity limit of EGRET has been included in the
GLF likelihood analysis. The EGRET sensitivity to a point source
changes depending on the location in the sky, and we used sensitivity
limit inferred from the signal-to-noise ratios of the EGRET sources as
a function of the Galactic latitude, as in NT06. The sensitivity limit
has a $\sim$30\% scatter even after binned by the Galactic
latitude. When the sensitivity limit is changed by $\pm$30\%, we find
that the GLF normalization and EGRB flux changes by $\pm
\sim$25\%. Since the faint-end slopes of the $(q,\gamma_1)$ GLF models
are $\gamma_1 \sim 1$, the faint-end cut-off luminosity $L_{\gamma,
  \min}$ of blazar GLF could also be important. We therefore repeated
the calculation with $L_{\gamma, \min}$ changed by a factor of 10
from the baseline model, but we find that the EGRB flux changes by
only $\sim$0.8\% and $\sim$1.7\% for U03($q$, $\gamma_1$) and H05($q$,
$\gamma_1$), respectively.  Therefore, our result is not sensitive to
$L_{\gamma, \min}$.

Next we examine the possible systematic uncertainties in the
observational determination of EGRB. The correct modeling of the
foreground, i.e., the Galactic diffuse emission is critical for the
EGRB measurement. However, there is a well-known problem of ``GeV
anomaly'', which is an excess of the observed diffuse flux compared
with the standard theoretical model of the Galactic diffuse emission
({Pohl} \& {Esposito} 1998; {Strong} {et~al.} 2004b; {de Boer}
{et~al.} 2005; {Kamae}, {Abe}, \& {Koi} 2005; {Stecker}, {Hunter}, \&
{Kniffen} 2008). The difference of the EGRB data of Strong et
al. (2004a) from the original EGRET data (Sreekumar et al. 1998) is a
result of modifying the model of the Galactic diffuse emission to
resolve the GeV anomaly, demonstrating that theoretical uncertainties
in the Galactic diffuse emission could have a significant effect on
the EGRB measurements. On the other hand, Stecker et al. (2008)
suggested that the most likely cause of the GeV anomaly is a
systematic error in the calibration of the EGRET detector at photon
energies beyond 1 GeV, and they derived the correction factors for the
flux measured by EGRET.  If their claim is correct, the correction
factors should be multiplied to the original EGRET measurements of
EGRB (Sreekumar et al. 1998), and such corrected data are shown as
S98+S08 in Fig. \ref{fig.egrb}.  Although
the overall corrected EGRB flux is still higher than the model
predictions, the corrected spectrum is similar to those of models. The
marked dip at $\sim$ GeV and hump at higher photon energies found in
the data of Strong et al. (2004a) are not seen.  It should also be
noted that the EGRET EGRB data at $\sim$50 MeV is considerably higher
than the neighboring COMPTEL data, again indicating some systematic
uncertainties in the EGRB flux estimates.

Based on these results and considerations, we conclude that most, and
probably all, of the EGRB flux can be explained by blazars and
non-blazar AGNs, with luminosity functions that are consistent with
the EGRET blazar data and the X-ray AGN surveys. We must await the
future {\it Fermi} data for a more robust conclusion about this issue.

\section{Predictions for the {\it Fermi} mission}
\label{section:Fermi}

\subsection{Expected Number of Blazars and Non-blazar AGNs}

\begin{figure*}
  \begin{center}
    \begin{tabular}{cc}
      \resizebox{90mm}{!}{\includegraphics{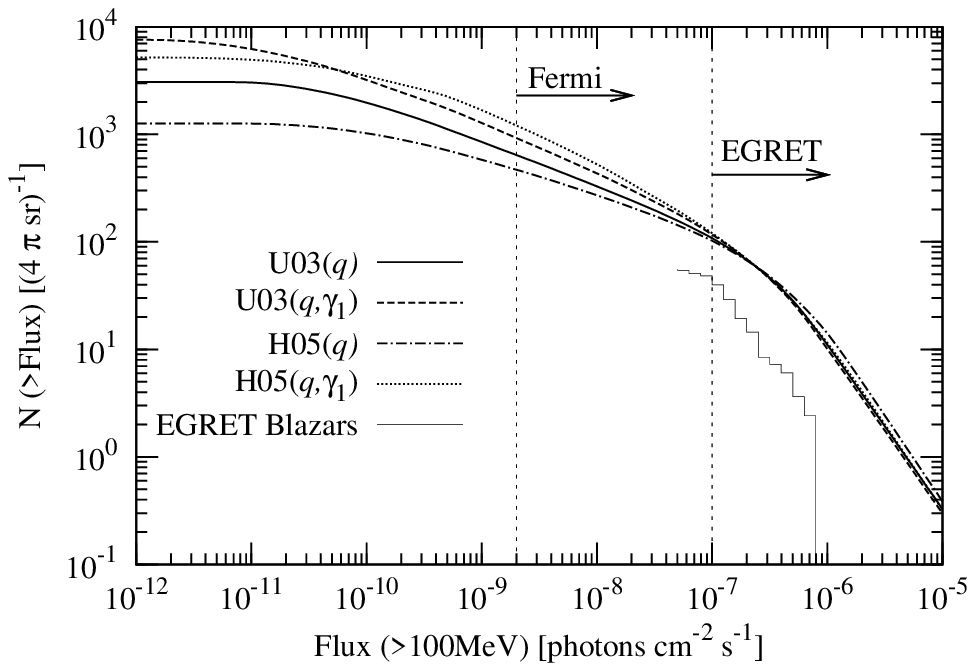}} &
      \resizebox{90mm}{!}{\includegraphics{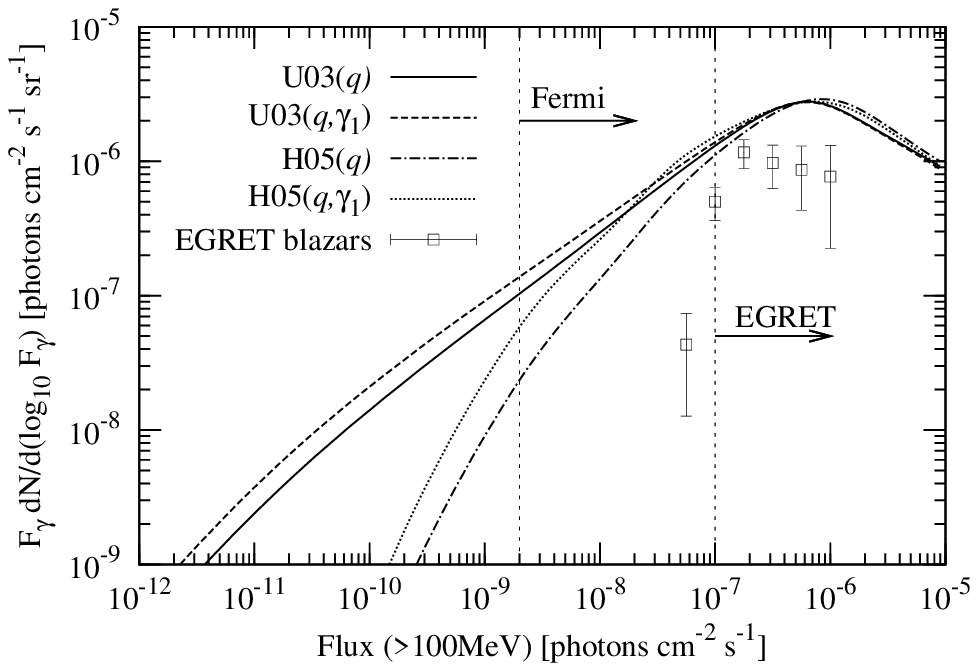}} \\
    \end{tabular}
    \caption{Left: the cumulative flux distribution of blazars. The
      four model curves correspond to the four different GLF models of
      U03($q$), H05($q$), U($q,\gamma_1$), and H05($q,\gamma_1$) are
      shown. The thin solid line shows the observed distribution of
      EGRET blazars. The detection limits of EGRET and {\it Fermi}
      are also shown. $Right$: the same as the left panel, but showing
      differential flux distribution multiplied by flux $F_\gamma$, to
      show the contribution to EGRB per logarithmic flux interval. The
      solid squares are the EGRET data with Poisson errors.}
    \label{fig.count}
  \end{center}
\end{figure*}
The left panel of Figure \ref{fig.count} shows the cumulative
distribution of $>$ 100 MeV photon flux of blazars.  The four GLF
models of U03($q$), U03($q,\gamma_1$), H05($q$), and
H05($q,\gamma_1$), predict that about 640, 930, 470, and 1200 blazars
should be detected by {\it Fermi}, respectively, where we assumed the
{\it Fermi} sensitivity to be $F_{\rm lim}=2\times 10^{-9}$ photons
cm$^{-2}$ s$^{-1}$ above 100 MeV.  This is a standard sensitivity
often used in the literature (e.g., Oh 2001; NT06; Venters \&
Pavlidou 2007), and close to $\sim 5 \sigma$ limit for a point source at high
Galactic latitude after a 1 year survey. See also the official Science
Requirement Document of {\it Fermi} available at
http://fermi.gsfc.nasa.gov/science/ for the {\it Fermi} performance.
Here, we calculated simply all the blazars without taking into account
the probability of identification by follow-up observations in other
wavebands. This is why the number of bright blazars is larger than the
EGRET observation in Fig. \ref{fig.count}; we have taken into account
the probability of blazar identification by detecting a radio
counterpart in the likelihood analysis of the EGRET blazars.  The
model of {Stecker} \& {Salamon} (1996), and the PLE and LDDE models in
{Narumoto} \& {Totani} (2006) predicted $\sim$10000, 5400, and 3000
blazars for the same sensitivity limit, respectively.  It is
remarkable that the GLF models in this work predict significantly
smaller numbers of blazars than previous studies. This is probably
because of the SED model newly used here; the IC component has its
broad SED peak at around 100 MeV, and the flux at lower and higher
energy band is relatively lower than the case of power-law SED,
because of the curvature of the spectrum.

The right panel of Figure \ref{fig.count} shows the differential flux
distribution of gamma-ray blazars multiplied by flux, showing the
contribution to the EGRB per unit logarithmic flux interval. From this
plot one can estimate how much fraction of EGRB will be resolved by
the {\it Fermi} mission.  The peak of the contribution to EGRB occurs at a
flux much brighter than the {\it Fermi} limit, meaning that EGRB from
blazars should practically be resolved into discrete sources.  We find
that the fraction of EGRB flux that should be resolved is 99, 98, 100,
and 100\% against the total blazar EGRB flux in the four GLF models of
U03($q$), U03($q,\gamma_1$), H05($q$), and H05($q,\gamma_1$),
respectively.  However, as seen in Fig. \ref{fig.egrb}, there is a
considerable contribution from non-blazar AGNs to the EGRB flux at 100
MeV, and our next interest is how much fraction of EGRB from
non-blazars will be resolved by {\it Fermi}.

\begin{figure*}
  \begin{center}
    \begin{tabular}{cc}
      \resizebox{90mm}{!}{\includegraphics{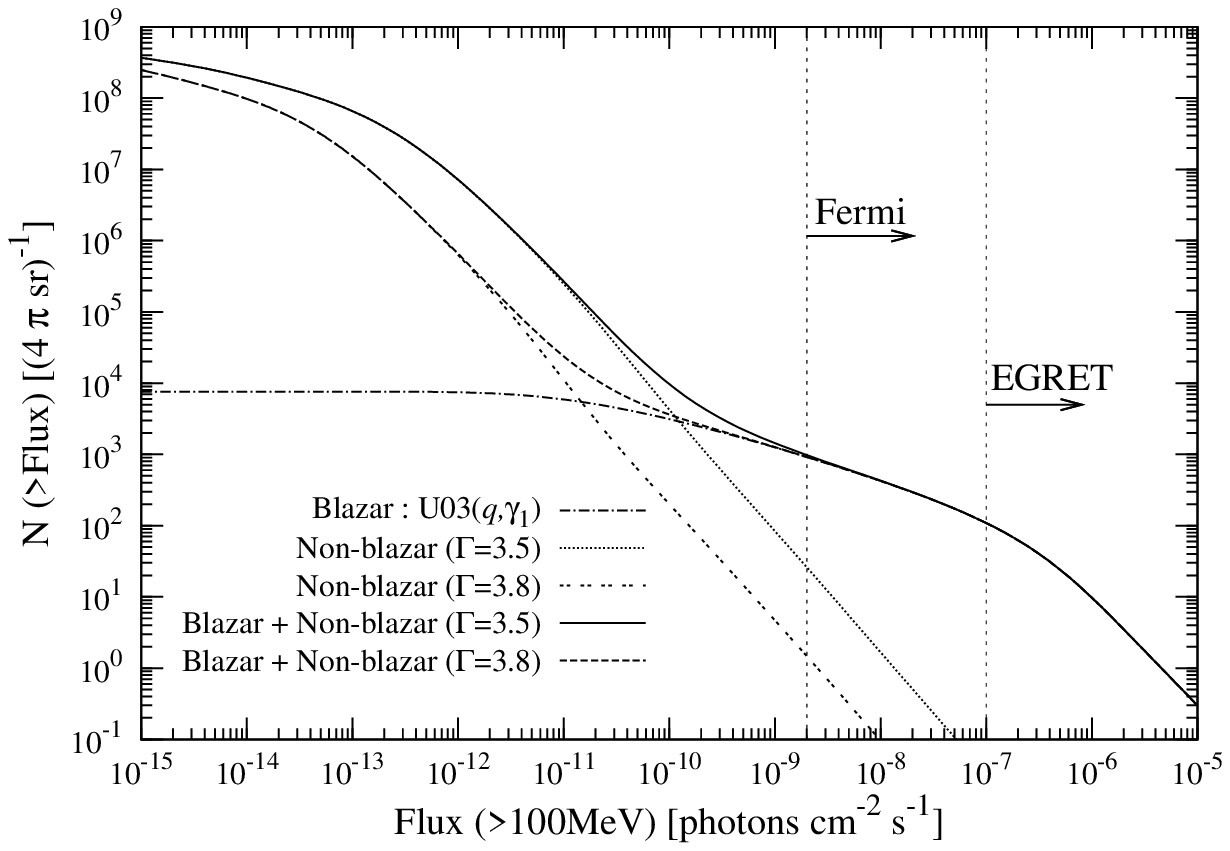}} &
      \resizebox{90mm}{!}{\includegraphics{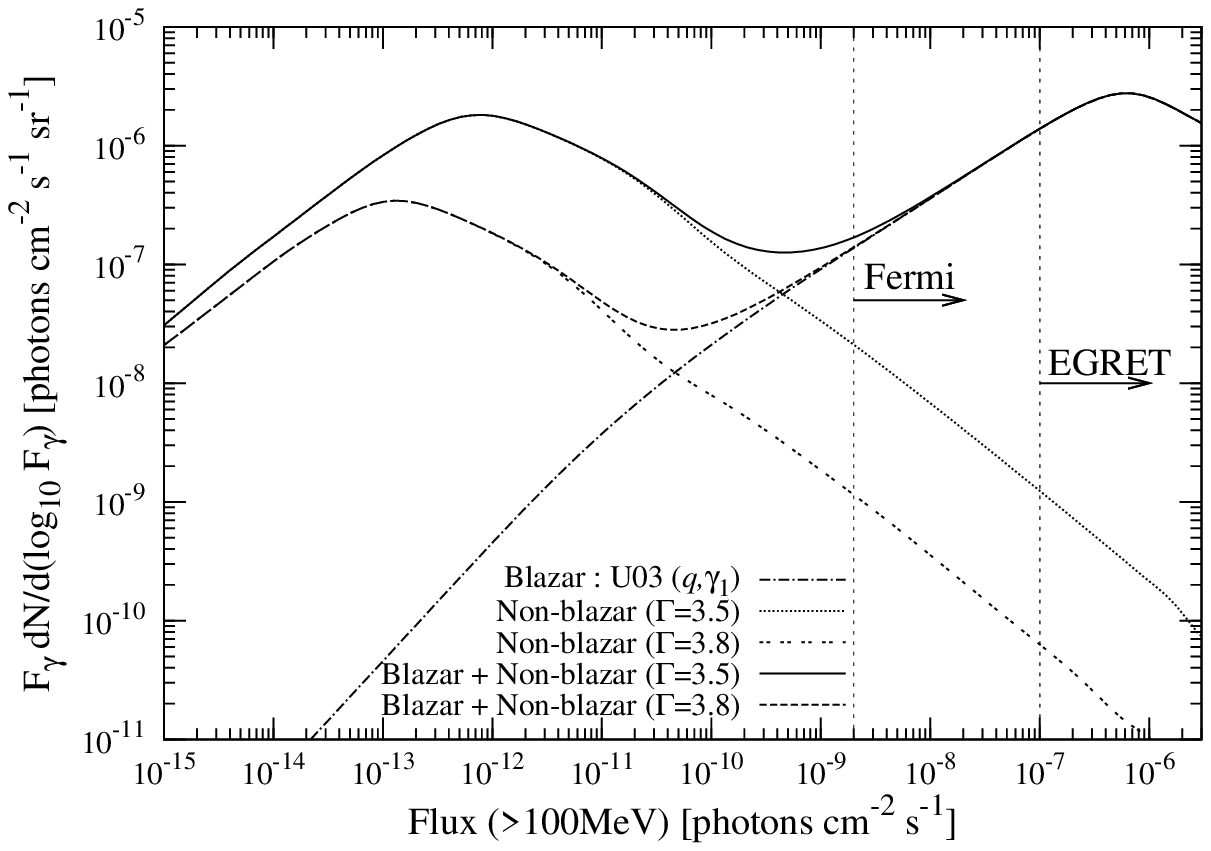}} \\
    \end{tabular}
    \caption{The same as Fig. \ref{fig.count}, but showing non-blazar AGNs
      as well, in addition to blazars.  The two models of non-blazar
      AGNs with different values of $\Gamma$ are shown. The total of
      blazar and non-blazar counts is also shown.}
  \label{fig.count_all}
  \end{center}
\end{figure*}

The left panel of Figure \ref{fig.count_all} shows the cumulative flux
distribution of blazars in the U03($q, \gamma_1$) model, non-blazar
AGNs, and the total of the two.  The expected number of non-blazar
AGNs detectable by EGRET is much smaller than unity in all sky, and it
is consistent with the fact that almost all extragalactic gamma-ray
sources detected by EGRET are blazars. However, about 1--30 non-blazar
AGNs would be detected by {\it Fermi}, by their soft nonthermal
emission from nonthermal electrons in coronae of accretion disks,
giving an interesting test for our model.  In fact, we can estimate
the $>$ 100 MeV gamma-ray flux from the observed hard X-ray flux of
NGC 4151 (the brightest Seyfert galaxy in all sky, {Sazonov} {et~al.}
2007) using the ITU08 model, and it becomes $\sim 3.3 \times 10^{-8}$
and $4.1 \times 10^{-9}$ photons cm$^{-2}$ s$^{-1}$ for $\Gamma = 3.5$
and 3.8, respectively. Therefore NGC 4151 is marginally detectable
when $\Gamma = 3.8$, and easily detectable with $\Gamma = 3.5$, which
are nicely consistent with the predicted source counts of
non-blazar AGNs detectable by {\it Fermi}.

On the other hand, the right panel of Figure \ref{fig.count_all}
indicates that EGRB from non-blazar AGNs is hardly resolved into
discrete sources by {\it Fermi}.  The predicted fraction of non-blazar
EGRB flux resolved by {\it Fermi} is 0.03 and 0.004 \% for the models
with $\Gamma=3.5$ and 3.8, respectively, against the total non-blazar
EGRB flux.  These results are in sharp contrast to those for blazars,
although the contributions to the total EGRB by the two populations
are comparable at around 100 MeV.  This is because of a large
difference between typical absolute luminosities of blazars and
non-blazar AGNs. The typical luminosity at 100 MeV of blazars and
non-blazars is $\nu L_\nu=10^{48.0}$ and $10^{42.5}$ erg/s, which are
corresponding to $L_X^*$ in the AGN XLF.  If the evolution is similar
for the two population, the characteristic redshift or distance to the
sources having main contribution to EGRB should be similar.
Therefore, a source population with smaller characteristic luminosity
is more difficult to resolve into discrete sources when the flux
sensitivity is fixed.

The above results mean that a considerable part of EGRB at around 100
MeV will remain unresolved even with the {\it Fermi} sensitivity,
because there is a considerable contribution from non-blazar AGNs to
EGRB at $\sim$100 MeV.  However, the contribution from non-blazars
should rapidly decrease with increasing photon energy, and almost all
of the total EGRB flux at $\gtrsim$ 1 GeV should be resolved into
discrete blazars by {\it Fermi}, if there is no significant source
contributing to EGRB other than blazars.  It should be noted that
these predictions can be tested by {\it Fermi} relatively easily,
without follow-up or cross check at other wavebands, once measurements
of source counts and EGRB flux have been done by the {\it Fermi} data.
Therefore this gives a simple and clear test for the theoretical model
presented here.

\subsection{Redshift and Luminosity Distribution}

\begin{figure*}
  \begin{center}
    \begin{tabular}{cc}
      \resizebox{90mm}{!}{\includegraphics{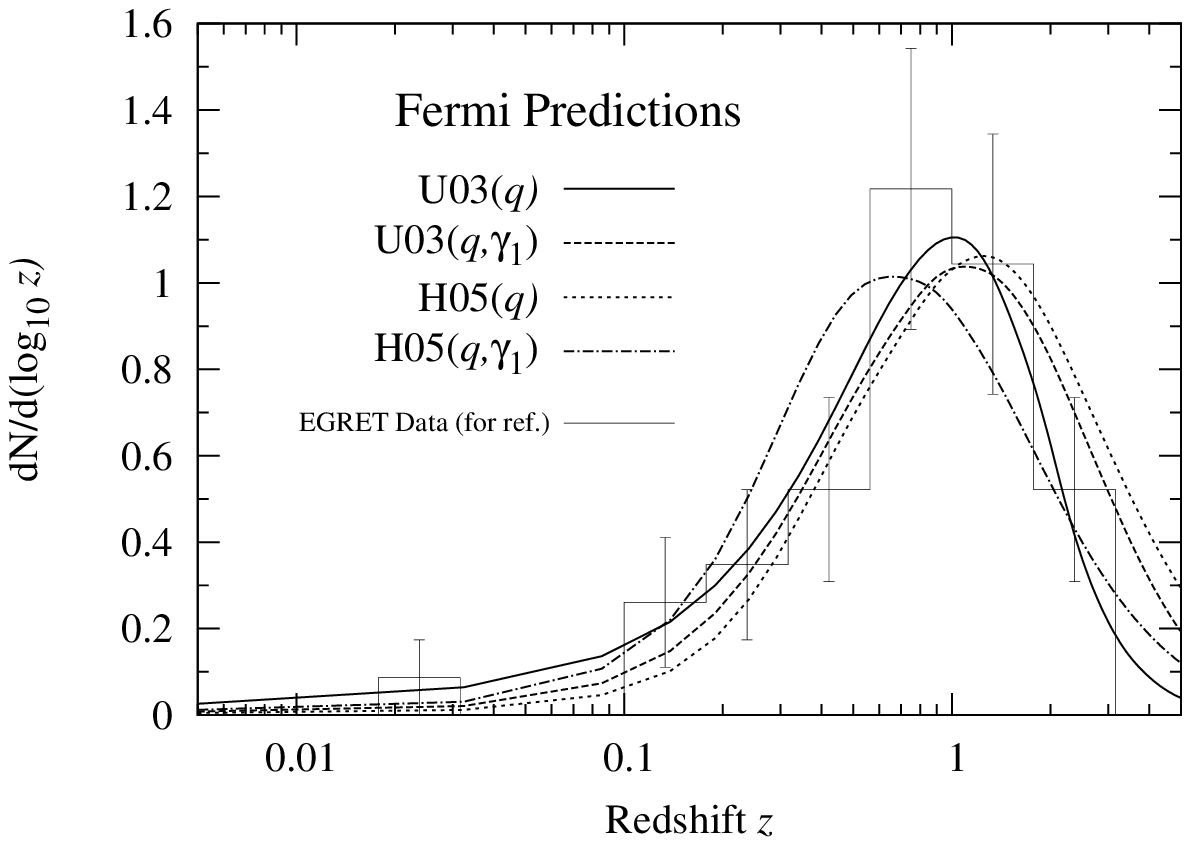}} &
      \resizebox{90mm}{!}{\includegraphics{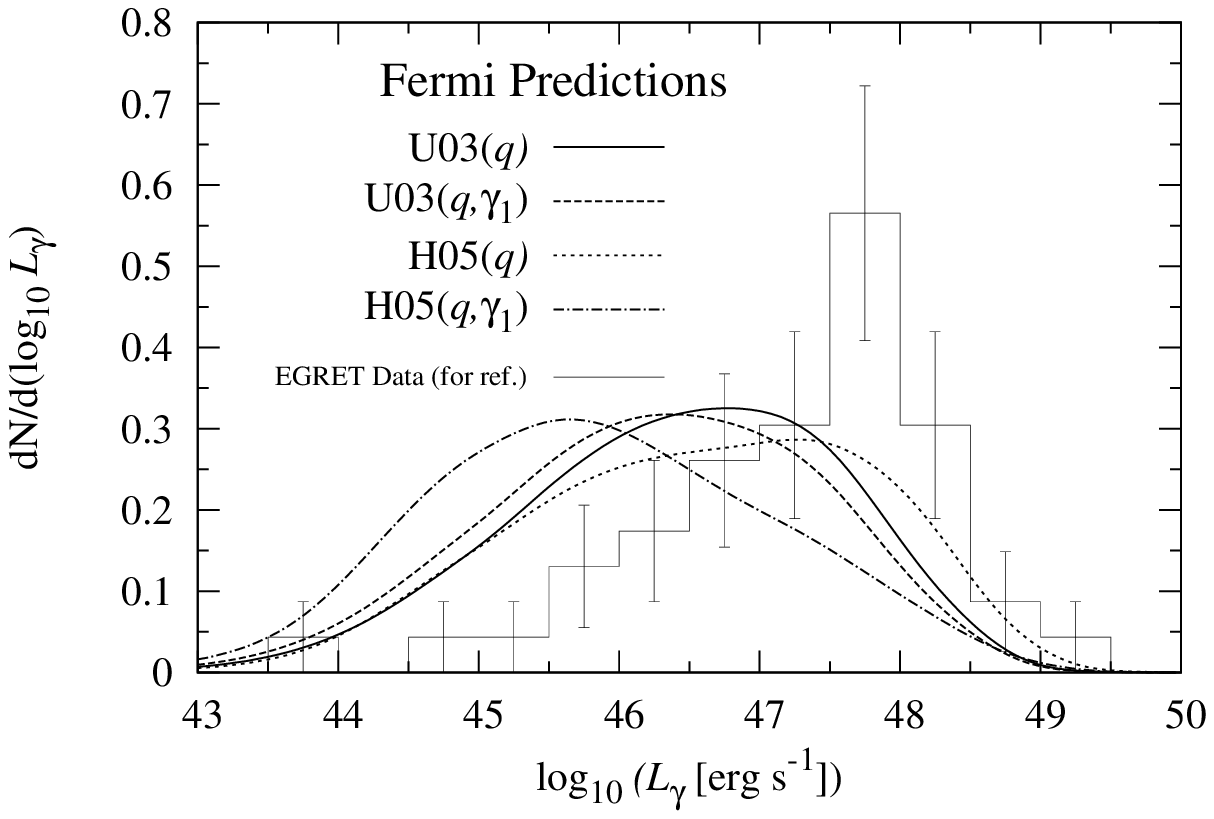}} \\ 
    \end{tabular}
    \caption{The predictions for the redshift and gamma-ray luminosity
      ($\nu L_\nu$ at rest frame 100 MeV) distributions of blazars for
      {\it Fermi}. The four model curves are for the four different
      GLF models of U03($q$), H05($q$), U($q,\gamma_1$), and
      H05($q,\gamma_1$).  The {\it Fermi} sensitivity limit is set as
      $F_{\rm lim}$ =$2 \times 10^{-9}$ photons cm$^{-2}$ s$^{-1}$ for
      photon flux above 100 MeV.  The observed distributions of EGRET
      blazars are shown, for comparison against the expected
      distributions of {\it Fermi} blazars.  
}
    \label{Fermi_dist}
  \end{center}
\end{figure*}
\label{z_L_Fermi}
Figure \ref{Fermi_dist} shows the redshift and luminosity distributions
of blazars that will be detected by {\it Fermi}, where we have again set the
{\it Fermi} sensitivity limit as $F_{\rm lim}=2\times 10^{-9}$ photons
cm$^{-2}$ s$^{-1}$ above 100 MeV. Since we normalized the total number
of detectable blazars, only the shapes of distribution should be
compared. 

{\it Fermi} will detect much fainter blazars than EGRET did. Since
$\gamma_1>1$, the H05($q,\gamma_1$) model predicts more faint blazars
than other models. The redshift distribution of EGRET blazars has
already extended to cosmologically significant range of $z \gtrsim 1$,
and the normalized distribution does not significantly change by the
{\it Fermi} sensitivity.  However, the absolute number of high-$z$
blazars will be increased, because the total number of blazars will
drastically be increased by {\it Fermi} from that of EGRET blazars.

There are some differences in the predictions by the four different
GLF models. Since the {\it Fermi} will dramatically increase the
statistics by a large number of detected blazars, a detailed
comparison between the GLF models and the {\it Fermi} data will
provide us a quantitative view of the evolutionary nature of blazar
GLF, once the redshifts of {\it Fermi} blazars are measured.  The
comparison between the blazar GLF and other AGN LFs in different
wavebands (e.g., X-ray, optical, and radio) will give us important
information about the jet activity evolution of AGNs, in comparison
with the mass accretion history onto SMBH that can be probed by disk
luminosity function.

\section{Discussion}
\label{section:discussion}

\begin{figure*}
  \begin{center}
\plotone{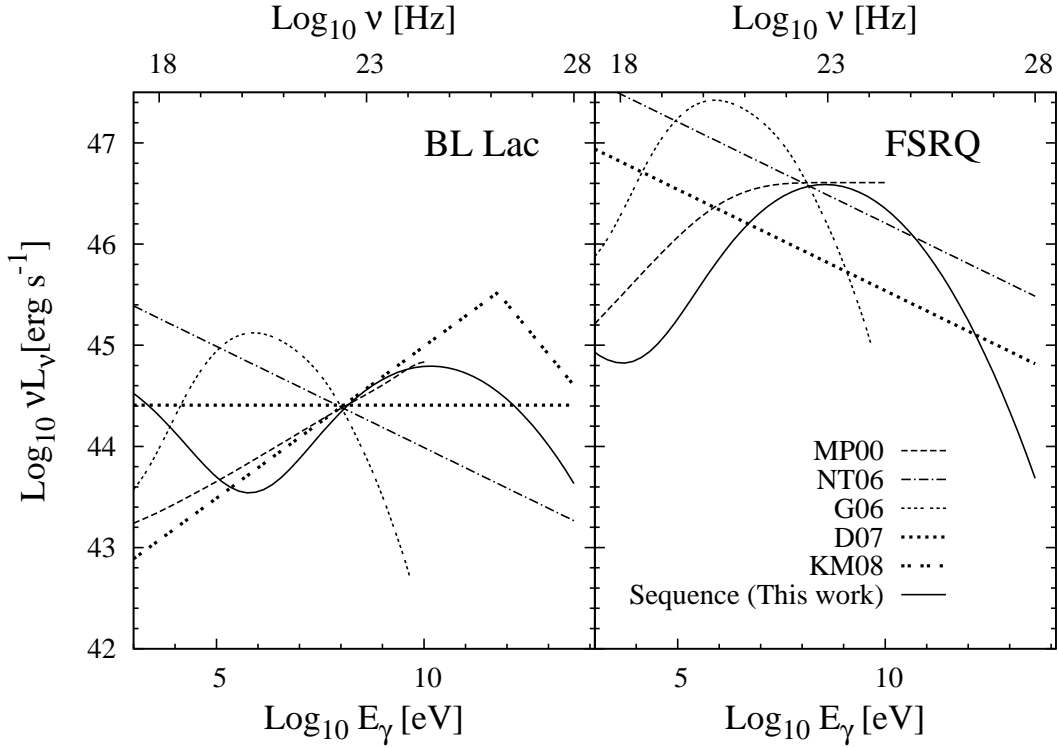}
\caption{The blazar SED models in our model in comparison with those
  in previous studies. BL Lacs and FSRQs are separately plotted in the
  left and right panels, respectively.  SEDs of our model are those
  having $L_{\gamma} = 10^{44.5}$ and $10^{46.6} \rm erg s^{-1}$ for
  BL Lac and FSRQ, respectively.  The MP00 curves are the EIC+SSC
  instantaneous injection model from their Fig 1. NT06 used a single
  power-law with the same index both for BL Lacs and FSRQs.  The G06
  curve (the same for BL Lac and FSRQ) is the SED with the synchrotron
  peak of $\nu_{\rm syn} = 10^{13.8}$ Hz used in their study.  D07
  used single power-law SEDs but with different indices for BL Lacs
  and FSRQs.  The KM08 curve is the case 3 SED used as a model of HBL
  SED.  The normalizations of these SEDs are from the original models
  for MP00 and D07, and otherwise they are fixed at $L_{\gamma} =
  10^{44.5}$ and $10^{46.6} \rm erg s^{-1}$ for BL Lac and FSRQ,
  respectively, where $L_{\gamma}$ is $\nu L_\nu$ at 100 MeV.  }
    \label{fig.sed_model}
  \end{center}
\end{figure*}

\subsection{Comparison with previous studies}
\label{subsection:past}
There are four important points in the modeling of the blazar
GLF. They are (1) the treatment of blazar population (e.g., BL
Lac/FSRQ separation), (2) modeling of blazar SED, (3) the functional
form of the GLF and its cosmological evolution, and (4) observational
inputs to determine the GLF parameters. Here we compare our model with
those of previous studies about these points.  Our results about the
expected number of Fermi blazars ($\sim$1000 blazars to the
sensitivity limit of $F_{\rm lim}=2\times 10^{-9}$ photons cm$^{-2}$
s$^{-1}$ above 100 MeV) and EGRB ($\sim 45$\% contribution by blazars
at 100 MeV) will also be compared with those of the previous studies.

{Stecker} \& {Salamon} (1996; hereafter SS96) treated the blazar as a
single population, with a single power-law SED.  They used the blazar
radio luminosity function of Dunlop \& Peacock (1990) with pure
luminosity evolution (PLE) as the blazar GLF model. The
gamma-ray/radio flux ratio was estimated by those of typical blazars,
and their blazar GLF was not statistically compared with the EGRET
blazar data of flux and redshift distribution. They found that blazars
can explain 100\% of EGRB, and $\sim$10000 blazars are expected to be
detected by {\it Fermi}.

{Chiang} \& {Mukherjee} (1998; hereafter CM98) treated the blazar as a
single population, with a single power-law SED.  CM98 assumed a single
power-law with a cut-off luminosity $L_{\min}$ for the luminosity
distribution of GLF, and adopted PLE for evolution.  In such a
modeling, the faint-end slope is difficult to estimate accurately, and
hence the uncertainty about EGRB flux becomes large.  They constrained
the blazar GLF parameters from a statistical comparison with the EGRET
blazar data.  They found that blazars can explain only 25\% of EGRB.

{M{\"u}cke} \& {Pohl} (2000; hereafter MP00) treated BL Lacs and FSRQs
separately. They used a theoretical model of the blazar SED of
{Dermer} \& {Schlickeiser} (1993), and their GLF model is described in
terms of the distribution of electron energy injection.  The energy
injection distribution is modeled according to the radio luminosity
function of Fanaroff-Riley (FR) I and II galaxies of Urry et
al. (1991) and Padovani \& Urry (1992) for BL Lacs and FSRQs,
respectively. These LFs assume PLE.  The GLF model parameters are then
determined by the flux and redshift distributions of EGRET blazars.
They found that blazars can explain 20-80\% of the EGRB depending on
the model parameters.

Narumoto \& Totani (2006; hereafter NT06) treated the blazar as a
single population, with a single power-law SED. They examined both the
PLE and LDDE model for GLE evolution, which are based on the radio LF
of blazars and XLF of AGNs, respectively.  The GLF model parameters
are determined by the fits to the flux and redshift distributions of
EGRET blazars. They found that LDDE model fits to EGRET data better
than the PLE model. They found that blazars can explain 25-50\% of the
EGRB, and $\sim$3000 blazars are expected to be detected by {\it
  Fermi}.

Giommi {et~al.} (2006; hereafter G06) estimated EGRB flux and spectrum
based on the radio luminosity function of blazars.  They first
calculated the cosmic radio background intensity, and then converted
it to the EGRB spectrum by using a typical SSC SED model. Therefore,
their model is not a GLF model for individual blazars, and not
compared statistically with the EGRET data of blazar flux and redshift
distribution.  Their model overproduced the observed EGRB spectrum.

Dermer (2007; hereafter D07) treated BL Lac and FSRQ separately, using
single power-law SEDs for the two populations with different spectral
indices. The variation of the absolute luminosity was not considered
within these populations, with the single luminosity of $L_\gamma =
2.5\times10^{44} \ \rm erg\ s^{-1}$ and $10^{46} \ \rm erg\ s^{-1}$
for BL Lacs and FSRQs, respectively.  Therefore, this is not a model
of GLF, but D07 rather considered various models of number density
evolution, and the model parameters of the number density evolution
are constrained by the redshift and flux distribution of EGRET
blazars.  They found that blazars can account for 12-19\% of EGRB at 1
GeV, and their model predicts $\sim$ 1000 blazars detectable by Fermi,
which is close to our result by a very different approach.

Kneiske \& Mannheim (2008; hereafter KM08) considered the contribution
to EBL from high energy cutoff BL Lacs (HBLs).  The spectrum was
assumed to be a broken power-law with the peak energy around 1 TeV for
HBLs. They employed the XLF of HBL which shows no evolution (Beckmann
et al. 2003), and a statistical comparison with the EGRET blazar data
was not made. They found that blazars can explain $\sim$80\% of EGRB by
including cascading emission (see also \S \ref{section:VHE}).

For comparison, we show the SEDs of our model corresponding to the
typical luminosities of BL Lacs ($L_{\gamma}\sim10^{44.5} \ \rm erg \
s^{-1}$) and FSRQs ($L_{\gamma}\sim10^{46.6} \ \rm erg \ s^{-1}$), as
well as those used in previous studies, in Fig. \ref{fig.sed_model}.

It should be noted that, in the studies that constrained the blazar
GLF parameters by a statistical comparison with the EGRET data, only
CM98, NT06, and this work take into account the probability of
identification of gamma-ray blazars by finding radio
counterparts. CM98 assumed that there is no correlation between
gamma-ray and radio luminosities of blazars. The assumption of no
correlation at all over a wide range of gamma-ray and radio
luminosities induces some inconsistencies (see discussion in Stecker
\& Salamon 2001), and it is physically reasonable to expect some level
of correlation between the two from the viewpoint of the standard
synchrotron and inverse Compton emission model of blazars.  In NT06
and this work, we introduced this correlation with a log-normal
scatter in the gamma/radio flux ratio, which is consistent with the
observation.

\subsection{Cascade of Very High Energy Gamma-Rays and EGRB}
\label{section:VHE}
KM08 estimated that HBL would account for $\sim$20\% of the EGRB by
cascading emission produced by absorption of very high energy
gamma-rays in IGM.  However, in our estimation the cascade emission
would not significantly contribute to EGRB (see
Fig.\ref{fig.cascade}), and the reason for this discrepancy is as
follows. In KM08, the minimum and maximum luminosities of HBL are set
as L$_{\rm min}=10^{43} \rm \ erg\ s^{-1}$ and L$_{\rm max}=10^{47}\rm
\ erg\ s^{-1}$, respectively (in $\nu L_\nu$ at 0.3 TeV).  Their SED
model is a broken-power law, which is independent of luminosity and
obtained from fitting to the TeV observations of Mkn421, PKS2005-489,
and PKS2155-304. However the luminosities of these three at 0.3 TeV
are about $10^{44}\rm \ erg\ s^{-1}$, which is relatively low in the
range of HBL luminosity assumed by KM08.  The use of the same SED up
to L$_{\rm max}=10^{47}\rm \ erg\ s^{-1}$ may overpredict the flux in
very high energy band, compared with the case of the SED sequence.
Therefore we conclude that the contribution to EGRB from the cascading
component is not significant provided that the SED sequence is valid.

\subsection{Effects of Flaring Activities}

It is well known that blazars show flaring activities. For example,
TeV Cherenkov telescopes H.E.S.S. and MAGIC have recently observed
minute-scale variability in TeV ranges ({Aharonian} {et~al.}  2007;
{Albert} {et~al.} 2007). The origin of such variability has been
discussed in many papers (Begelman et al. 2008; Ghisellini \&
Tavecchio 2008a; Ghisellini et al. 2008; Katarzy{\'n}ski et al. 2008;
Mastichiadis \& Moraitis 2008; Giannios et al. 2009), but it is still
rather uncertain.  Therefore it is difficult to model the flaring
activity quantitatively and incorporate in our analysis.  We used the
time-averaged gamma-ray flux of blazars in the third EGRET catalog,
which is the mean of several viewing periods during the total EGRET
observation period (see \S2.1 of NT06 for more details).  Typical
duration of one viewing period is 2--3 weeks, and flaring activities
within this scale have already been taken into account.  Flaring
activities that might have happened at epochs out of the viewing
periods are not taken into account, but still the flaring effect
should have been incorporated to some extent by the use of time
averaged flux. More observational information about blazar flares
(light curves and frequency, etc.) is required to examine this issue
in more detail, and future Fermi data will be useful.

\section{Conclusions}
\label{section:conclusions}

In this paper, we constructed a new model of the blazar gamma-ray
luminosity function (GLF), taking into account the blazar SED sequence
and the LDDE luminosity function inferred from X-ray observations of
AGNs. An implicit assumption here is that the jet activity of AGNs is
associated with the accretion activity, and hence the blazar
luminosity function has a similar evolution to that of X-ray AGNs.
The GLF model parameters are constrained by carefully fitting to the
observed flux and redshift distribution of the EGRET blazars. By this
model, for the first time, we can predict the spectrum of the
extragalactic gamma-ray background (EGRB) in a non-trivial way, rather
than assuming a simple functional form such as power-law spectra.

The absorption of gamma-rays in IGM by interaction with the cosmic
infrared background (CIB), and the reprocessed cascade emission from
electrons/positrons produced in IGM are also taken into account in the
EGRB calculation.  We found that, provided that the blazar SED
sequence is valid, the amount of EGRB energy flux absorbed by CIB
interaction (at photon energy $\gtrsim$ 10 GeV) is not large and the
EGRB spectrum at lower photon energy bands is not significantly
affected by the secondary cascade emission.

The contribution from non-blazar AGNs to EGRB is also considered, by
using the nonthermal coronal electron model of Inoue et al. (2008).
This model is a natural extension of the standard X-ray spectral model
of AGN emission from the disk and corona region, and predicts the
nonthermal emission extending from MeV to higher energy band with a
steep power-law. This model gives a natural explanation for the
observed cosmic MeV background, and we examined whether X-ray to GeV
gamma-ray background radiation can be accounted for by the two
population model including blazars and non-blazar AGNs.

Our model predicts that the EGRB flux from blazars at 100 MeV is only
about 45 \% of the observed EGRB flux of Strong et al. (2004a).
However, it is possible to account for more than 80\% of the observed
EGRB flux by the sum of blazars and non-blazar AGNs. The predicted
spectrum is also in good agreement with the observed data from X-ray
to $\sim$1 GeV, indicating that the EGRB below 1 GeV can mostly be
explained by our two population model. The two components have a
comparable contribution to EGRB at $\sim$100 MeV, and the blazar
component is dominant at the higher energy range while the non-blazar
component at the lower. Therefore, the higher EGRB flux at 100 MeV
found by this work than many of previous studies is not by change in
the EGRB flux from blazars, but by adding non-blazar AGNs.

The EGRB spectrum of Strong et al. (2004a) shows a hump beyond 3 GeV,
which cannot be explained by our model.  The EGRB flux determined
originally by the EGRET team (Sreekumar et al. 1998) is higher than
our model, and the observed spectrum is harder.  However, measurements
of EGRB suffer from the uncertainties in the theoretical modeling of
the foreground diffuse Galactic emission, and perhaps from systematic
errors in the calibration of the EGRET detector above 1 GeV (Stecker
et al. 2008). We must await the {\it Fermi} data to examine these
issues and whether an additional, completely different source is
necessary to explain all of EGRB.  We conclude that most (at least
more than 50\%), and probably all, of the observed high-energy
background radiation from X-ray to 1 GeV spanning 6 orders of
magnitude in photon energy can be accounted for by AGNs including
blazars.

We predicted the flux distribution of blazars by our model, and we
found that 600--1200 blazars in all sky should be detected by {\it
  Fermi}, assuming a sensitivity limit of $F_{\rm lim}=2\times
10^{-9}$ photons cm$^{-2}$ s$^{-1}$ above 100 MeV. This number is
significantly lower than the predictions by many of the previous
studies.  The expected number of the non-blazar AGNs at 100 MeV band
is much smaller than unity in all sky at the EGRET sensitivity, but
about 1--30 of such population are expected be detected by {\it
  Fermi}.  {\it Fermi} should resolve almost all of the EGRB flux from
blazars into discrete blazars, with percentages of $\gtrsim$99\%.  On
the other hand, less than 0.1\% of the EGRB flux from non-blazar AGNs
can be resolved into discrete sources.  Therefore, we have a clear
prediction: {\it Fermi} will resolve almost all of the EGRB flux into
discrete sources at photon energies $\gtrsim 1$ GeV where blazars are
dominant, while a significant fraction of the EGRB flux will remain
unresolved in the low energy band of $\lesssim$ 100 MeV where
non-blazar AGNs have a significant contribution.  This prediction can
easily be tested, only with the source counts and the EGRB estimates
by {\it Fermi} data.

We also predicted the redshift and absolute luminosity distributions
for the {\it Fermi} blazars. Future {\it Fermi} data set with measured
redshifts will enable us to get a quantitative measurement of the
blazar GLF and its evolution with a much larger statistics than
EGRET. A direct comparison between blazar GLF and X-ray AGN LF could
be done, making it possible to discuss the relation between the cosmic
histories of jet activity and accretion activity of AGNs. Our model
will be able to serve as a guide in such studies.  For example, if the
jet luminosity is proportional to the mass accretion rate, we expect a
larger jet/disk ratio than assumed in this work for AGNs with low
Eddington ratio, because X-ray luminosity scales as $\propto \dot m^2$
in the RIAF regime.  Then, we may expect a different behavior of
blazar GLF from that of AGN XLF at faint luminosity range. Such an
evidence is already seen from our fit to the EGRET data (see
Fig. \ref{EGRET_dist}), and it will be tested more quantitatively by
the {\it Fermi} data soon.

\acknowledgments

We thank Takuro Narumoto and Tuneyoshi Kamae for their help and useful
discussions.  This work was supported by the Grant-in-Aid for the
Global COE Program "The Next Generation of Physics, Spun from
Universality and Emergence" and Scientific Research (19047003,
19740099) from the Ministry of Education, Culture, Sports, Science and
Technology (MEXT) of Japan.

\bibliography{}


\begin{thebibliography}{}


\bibitem[{Aharonian}, {Coppi}, \&  {Voelk} 1994]{1994ApJ...423L...5A}
{Aharonian}, F.~A., {Coppi}, P.~S., \& {Voelk}, H.~J. 1994, \apjl, 423, L5

\bibitem[{Aharonian}, {Akhperjanian}, {Bazer-Bachi},  {Behera}, {Beilicke}, {Benbow}, {Berge}, {Bernl{\"o}hr}, {Boisson}, {Bolz},  {Borrel}, {Boutelier}, {Braun}, {Brion}, {Brown}, {B{\"u}hler},  {B{\"u}sching}, {Bulik}, {Carrigan}, {Chadwick}, {Clapson}, {Chounet},  {Coignet}, {Cornils}, {Costamante}, {Degrange}, {Dickinson},  {Djannati-Ata{\"i}}, {Domainko}, {Drury}, {Dubus}, {Dyks}, {Egberts},  {Emmanoulopoulos}, {Espigat}, {Farnier}, {Feinstein}, {Fiasson},  {F{\"o}rster}, {Fontaine}, {Funk}, {Funk}, {F{\"u}{\ss}ling}, {Gallant},  {Giebels}, {Glicenstein}, {Gl{\"u}ck}, {Goret}, {Hadjichristidis}, {Hauser},  {Hauser}, {Heinzelmann}, {Henri}, {Hermann}, {Hinton}, {Hoffmann}, {Hofmann},  {Holleran}, {Hoppe}, {Horns}, {Jacholkowska}, {de Jager}, {Kendziorra},  {Kerschhaggl}, {Kh{\'e}lifi}, {Komin}, {Kosack}, {Lamanna}, {Latham}, {Le  Gallou}, {Lemi{\`e}re}, {Lemoine-Goumard}, {Lenain}, {Lohse}, {Martin},  {Martineau-Huynh}, {Marcowith}, {Masterson}, {Maurin}, {McComb}, {Moderski},  {Moulin}, {de Naurois}, {Nedbal}, {Nolan}, {Olive}, {Orford}, {Osborne},  {Ostrowski}, {Panter}, {Pedaletti}, {Pelletier}, {Petrucci}, {Pita},  {P{\"u}hlhofer}, {Punch}, {Ranchon}, {Raubenheimer}, {Raue}, {Rayner},  {Renaud}, {Ripken}, {Rob}, {Rolland}, {Rosier-Lees}, {Rowell}, {Rudak},  {Ruppel}, {Sahakian}, {Santangelo}, {Saug{\'e}}, {Schlenker}, {Schlickeiser},  {Schr{\"o}der}, {Schwanke}, {Schwarzburg}, {Schwemmer}, {Shalchi}, {Sol},  {Spangler}, {Stawarz}, {Steenkamp}, {Stegmann}, {Superina}, {Tam},  {Tavernet}, {Terrier}, {van Eldik}, {Vasileiadis}, {Venter}, {Vialle},  {Vincent}, {Vivier}, {V{\"o}lk}, {Volpe}, {Wagner}, {Ward}, \&  {Zdziarski} 2007]{2007ApJ...664L..71A}
{Aharonian}, F., {et al.} 2007, \apjl, 664, L71

\bibitem[{Albert}, {Aliu}, {Anderhub}, {Antoranz},  {Armada}, {Baixeras}, {Barrio}, {Bartko}, {Bastieri}, {Becker}, {Bednarek},  {Berger}, {Bigongiari}, {Biland}, {Bock}, {Bordas}, {Bosch-Ramon}, {Bretz},  {Britvitch}, {Camara}, {Carmona}, {Chilingarian}, {Coarasa}, {Commichau},  {Contreras}, {Cortina}, {Costado}, {Curtef}, {Danielyan}, {Dazzi}, {De  Angelis}, {Delgado}, {de los Reyes}, {De Lotto}, {Domingo-Santamar{\'{\i}}a},  {Dorner}, {Doro}, {Errando}, {Fagiolini}, {Ferenc}, {Fern{\'a}ndez}, {Firpo},  {Flix}, {Fonseca}, {Font}, {Fuchs}, {Galante}, {Garc{\'{\i}}a-L{\'o}pez},  {Garczarczyk}, {Gaug}, {Giller}, {Goebel}, {Hakobyan}, {Hayashida},  {Hengstebeck}, {Herrero}, {H{\"o}hne}, {Hose}, {Hrupec}, {Hsu}, {Jacon},  {Jogler}, {Kosyra}, {Kranich}, {Kritzer}, {Laille}, {Lindfors}, {Lombardi},  {Longo}, {L{\'o}pez}, {L{\'o}pez}, {Lorenz}, {Majumdar}, {Maneva},  {Mannheim}, {Mansutti}, {Mariotti}, {Mart{\'{\i}}nez}, {Mazin}, {Merck},  {Meucci}, {Meyer}, {Miranda}, {Mirzoyan}, {Mizobuchi}, {Moralejo}, {Nieto},  {Nilsson}, {Ninkovic}, {O{\~n}a-Wilhelmi}, {Otte}, {Oya}, {Paneque},  {Panniello}, {Paoletti}, {Paredes}, {Pasanen}, {Pascoli}, {Pauss}, {Pegna},  {Persic}, {Peruzzo}, {Piccioli}, {Prandini}, {Puchades}, {Raymers}, {Rhode},  {Rib{\'o}}, {Rico}, {Rissi}, {Robert}, {R{\"u}gamer}, {Saggion}, {Saito},  {S{\'a}nchez}, {Sartori}, {Scalzotto}, {Scapin}, {Schmitt}, {Schweizer},  {Shayduk}, {Shinozaki}, {Shore}, {Sidro}, {Sillanp{\"a}{\"a}}, {Sobczynska},  {Stamerra}, {Stark}, {Takalo}, {Tavecchio}, {Temnikov}, {Tescaro}, {Teshima},  {Torres}, {Turini}, {Vankov}, {Vitale}, {Wagner}, {Wibig}, {Wittek},  {Zandanel}, {Zanin}, \& {Zapatero} 2007]{2007ApJ...669..862A}
{Albert}, J., {et al.} 2007, \apj, 669, 862

\bibitem[Ajello et al.(2008)]{2008arXiv0808.3377A} Ajello, M., et al.\ 
2008, ArXiv e-prints, 808, arXiv:0808.3377 

\bibitem[Beckmann et 
al.(2003)]{2003A&A...401..927B} Beckmann, V., Engels, D., Bade, N., \& Wucknitz, O.\ 2003, \aap, 401, 927 

\bibitem[Begelman et al.(2008)]{2008MNRAS.384L..19B} Begelman, M.~C., 
Fabian, A.~C., \& Rees, M.~J.\ 2008, \mnras, 384, L19 

\bibitem[Bhattacharya et al.(2008)]{2008arXiv0811.4388B} Bhattacharya, D., 
Sreekumar, P., \& Mukherjee, R.\ 2008, arXiv:0811.4388 

\bibitem[{Chiang}, {Fichtel}, {von Montigny}, {Nolan},  \& {Petrosian} 1995]{1995ApJ...452..156C}
{Chiang}, J., {Fichtel}, C.~E., {von Montigny}, C., {Nolan}, P.~L., \&  {Petrosian}, V. 1995, \apj, 452, 156

\bibitem[{Chiang} \& {Mukherjee} 1998]{1998ApJ...496..752C}
{Chiang}, J. \& {Mukherjee}, R. 1998, \apj, 496, 752 (CM98)

\bibitem[{de Boer}, {Sander}, {Zhukov}, {Gladyshev},  \& {Kazakov} 2005]{2005A&A...444...51D}
{de Boer}, W., {Sander}, C., {Zhukov}, V., {Gladyshev}, A.~V., \& {Kazakov},  D.~I. 2005, \aap, 444, 51

\bibitem[{Dermer} 2007]{2007ApJ...659..958D}
{Dermer}, C.~D. 2007, \apj, 659, 958 (D07)

\bibitem[{Dermer} \& {Schlickeiser} 1993]{1993ApJ...416..458D}
{Dermer}, C.~D. \& {Schlickeiser}, R. 1993, \apj, 416, 458

\bibitem[{Donato}, {Ghisellini}, {Tagliaferri}, \&  {Fossati} 2001]{2001A&A...375..739D}
{Donato}, D., {Ghisellini}, G., {Tagliaferri}, G., \& {Fossati}, G. 2001, \aap,  375, 739 (D01)

\bibitem[Falcke \& Biermann(1995)]{1995A&A...293..665F} Falcke, H., \& Biermann, P.~L.\ 1995, \aap, 293, 665 

\bibitem[Falcke et al.(1995)]{1995A&A...298..375F} Falcke, H., Malkan, M.~A., \& Biermann, P.~L.\ 1995, \aap, 298, 375 


\bibitem[Falcke et al.(2004)]{2004A&A...414..895F} Falcke, H., K{\"o}rding, E., \& Markoff, S.\ 2004, \aap, 414, 895 

\bibitem[{Fan}, {Dai}, \& {Wei} 2004]{2004A&A...415..483F}
{Fan}, Y.~Z., {Dai}, Z.~G., \& {Wei}, D.~M. 2004, \aap, 415, 483


\bibitem[Abdo(2009a)]{2009arXiv0902.1340A} Fermi/LAT Collaboration: A.~A.~Abdo {et al.} \ 2009a, 
arXiv:0902.1340 

\bibitem[Abdo(2009b)]{2009arXiv0902.1559A} Fermi/LAT Collaboration: A.~A.~Abdo {et al.} \ 2009b, 
arXiv:0902.1559 

\bibitem[Dunlop 
\& Peacock(1990)]{1990MNRAS.247...19D} Dunlop, J.~S., \& Peacock, J.~A.\ 1990, \mnras, 247, 19 

\bibitem[Fermi/LAT Collaboration (2009)]{2009arXiv0902.1089F} 
Fermi/LAT Collaboration: W.~B.~Atwood {et al.} 2009, arXiv:0902.1089 

\bibitem[{Fichtel}, {Simpson}, \&  {Thompson} 1978]{1978ApJ...222..833F}
{Fichtel}, C.~E., {Simpson}, G.~A., \& {Thompson}, D.~J. 1978, \apj, 222, 833

\bibitem[{Fossati}, {Celotti}, {Ghisellini}, \&  {Maraschi} 1997]{fossati-97}
{Fossati}, G., {Celotti}, A., {Ghisellini}, G., \& {Maraschi}, L. 1997, \mnras,  289, 136

\bibitem[{Fossati}, {Maraschi}, {Celotti}, {Comastri},  \& {Ghisellini} 1998]{fossati-98}
{Fossati}, G., {Maraschi}, L., {Celotti}, A., {Comastri}, A., \& {Ghisellini},  G. 1998, \mnras, 299, 433

\bibitem[Gallo et al.(2005)]{2005Natur.436..819G} Gallo, E., Fender, R., 
Kaiser, C., Russell, D., Morganti, R., Oosterloo, T., 
\& Heinz, S.\ 2005, \nat, 436, 819 

\bibitem[Gallo et al.(2003)]{2003MNRAS.344...60G} Gallo, E., Fender, R.~P., 
\& Pooley, G.~G.\ 2003, \mnras, 344, 60 

\bibitem[Ghisellini et al.(1998)]{1998MNRAS.301..451G} Ghisellini, G., 
Celotti, A., Fossati, G., Maraschi, L., 
\& Comastri, A.\ 1998, \mnras, 301, 451 

\bibitem[Ghisellini et al. (2009)]{2009arXiv0903.2043G} Ghisellini, G., 
Maraschi, L., \& Tavecchio, F.\ 2009, arXiv:0903.2043 

\bibitem[Ghisellini 
\& Tavecchio(2008a)]{2008MNRAS.386L..28G} Ghisellini, G., \& Tavecchio, F.\ 2008, \mnras, 386, L28 

\bibitem[Ghisellini \& Tavecchio(2008b)]{2008MNRAS.387.1669G} 
Ghisellini, G., \& Tavecchio, F.\ 2008, \mnras, 387, 1669 

\bibitem[Ghisellini et al.(2008)]{2008MNRAS.tmpL.146G} Ghisellini, G., 
Tavecchio, F., Bodo, G., \& Celotti, A.\ 2008, \mnras, L146 

\bibitem[Giannios et al.(2009)]{2009arXiv0901.1877G} Giannios, D., 
Uzdensky, D.~A., \& Begelman, M.~C.\ 2009, arXiv:0901.1877 

\bibitem[{Gilli}, {Comastri}, \&  {Hasinger} 2007]{2007A&A...463...79G}
{Gilli}, R., {Comastri}, A., \& {Hasinger}, G. 2007, \aap, 463, 79

\bibitem[{Giommi}, {Colafrancesco}, {Cavazzuti},  {Perri}, \& {Pittori} 2006]{2006A&A...445..843G}
{Giommi}, P., {Colafrancesco}, S., {Cavazzuti}, E., {Perri}, M., \& {Pittori},  C. 2006, \aap, 445, 843 (G06)

\bibitem[{Gruber}, {Matteson}, {Peterson}, \&  {Jung} 1999]{1999ApJ...520..124G}
{Gruber}, D.~E., {Matteson}, J.~L., {Peterson}, L.~E., \& {Jung}, G.~V. 1999,  \apj, 520, 124

\bibitem[{Hasinger}, {Miyaji}, \&  {Schmidt} 2005]{2005A&A...441..417H}
{Hasinger}, G., {Miyaji}, T., \& {Schmidt}, M. 2005, \aap, 441, 417 (H05)

\bibitem[{Inoue}, {Totani}, \&  {Ueda} 2008]{2008ApJ...672L...5I}
{Inoue}, Y., {Totani}, T., \& {Ueda}, Y. 2008, \apjl, 672, L5 (ITU08)

\bibitem[{Kamae}, {Abe}, \& {Koi} 2005]{2005ApJ...620..244K}
{Kamae}, T., {Abe}, T., \& {Koi}, T. 2005, \apj, 620, 244

\bibitem[{Kamae}, {Karlsson}, {Mizuno}, {Abe}, \&  {Koi} 2006]{2006ApJ...647..692K}
{Kamae}, T., {Karlsson}, N., {Mizuno}, T., {Abe}, T., \& {Koi}, T. 2006, \apj,  647, 692

\bibitem[{Kappadath}, {Ryan}, {Bennett}, {Bloemen},  {Forrest}, {Hermsen}, {Kippen}, {McConnell}, {Schoenfelder}, {van Dijk},  {Varendorff}, {Weidenspointner}, \& {Winkler} 1996]{1996A&AS..120C.619K}
{Kappadath}, S.~C., {et al.} 1996,  \aaps, 120, C619+

\bibitem[Katarzy{\'n}ski et al.(2008)]{2008MNRAS.390..371K} 
Katarzy{\'n}ski, K., Lenain, J.-P., Zech, A., Boisson, C., 
\& Sol, H.\ 2008, \mnras, 390, 371 

\bibitem[{Kato}, {Fukue}, \&  {Mineshige} 1998]{1998bhad.conf.....K}
{Kato}, S., {Fukue}, J., \& {Mineshige}, S., eds. 1998, {Black-hole accretion  disks}

\bibitem[{Keshet}, {Waxman}, \&  {Loeb} 2004]{2004JCAP...04..006K}
{Keshet}, U., {Waxman}, E., \& {Loeb}, A. 2004, Journal of Cosmology and  Astro-Particle Physics, 4, 6

\bibitem[{Kneiske}, {Bretz}, {Mannheim}, \&  {Hartmann} 2004]{2004A&A...413..807K}
{Kneiske}, T.~M., {Bretz}, T., {Mannheim}, K., \& {Hartmann}, D.~H. 2004, \aap,  413, 807

\bibitem[{Kneiske} \& {Mannheim} 2008]{2008A&A...479...41K}
{Kneiske}, T.~M. \& {Mannheim}, K. 2008, \aap, 479, 41 (KM08)

\bibitem[K{\"o}rding et al.(2006)]{2006MNRAS.369.1451K} K{\"o}rding, E.~G., 
Fender, R.~P., \& Migliari, S.\ 2006, \mnras, 369, 1451
 
\bibitem[Kubo et al.(1998)]{1998ApJ...504..693K} Kubo, H., Takahashi, T., 
Madejski, G., Tashiro, M., Makino, F., Inoue, S., \& Takahara, F.\ 1998, \apj, 504, 693

\bibitem[{Laor} \& {Behar} 2008]{2008arXiv0808.0637L}
{Laor}, A. \& {Behar}, E. 2008, ArXiv e-prints, 808

\bibitem[{MAGIC Collaboration}, {Albert},  {Aliu}, {Anderhub}, {Antonelli}, {Antoranz}, {Backes}, {Baixeras}, {Barrio},  {Bartko}, {Bastieri}, {Becker}, {Bednarek}, {Berger}, {Bernardini},  {Bigongiari}, {Biland}, {Bock}, {Bonnoli}, {Bordas}, {Bosch-Ramon}, {Bretz},  {Britvitch}, {Camara}, {Carmona}, {Chilingarian}, {Commichau}, {Contreras},  {Cortina}, {Costado}, {Covino}, {Curtef}, {Dazzi}, {De Angelis}, {Cea del  Pozo}, {de los Reyes}, {De Lotto}, {De Maria}, {De Sabata}, {Mendez},  {Dominguez}, {Dorner}, {Doro}, {Errando}, {Fagiolini}, {Ferenc},  {Fern{\'a}ndez}, {Firpo}, {Fonseca}, {Font}, {Galante}, {L{\'o}pez},  {Garczarczyk}, {Gaug}, {Goebel}, {Hayashida}, {Herrero}, {H{\"o}hne}, {Hose},  {Hsu}, {Huber}, {Jogler}, {Kneiske}, {Kranich}, {La Barbera}, {Laille},  {Leonardo}, {Lindfors}, {Lombardi}, {Longo}, {L{\'o}pez}, {Lorenz},  {Majumdar}, {Maneva}, {Mankuzhiyil}, {Mannheim}, {Maraschi}, {Mariotti},  {Mart{\'{\i}}nez}, {Mazin}, {Meucci}, {Meyer}, {Miranda}, {Mirzoyan},  {Mizobuchi}, {Moles}, {Moralejo}, {Nieto}, {Nilsson}, {Ninkovic}, {Otte},  {Oya}, {Panniello}, {Paoletti}, {Paredes}, {Pasanen}, {Pascoli}, {Pauss},  {Pegna}, {Perez-Torres}, {Persic}, {Peruzzo}, {Piccioli}, {Prada},  {Prandini}, {Puchades}, {Raymers}, {Rhode}, {Rib{\'o}}, {Rico}, {Rissi},  {Robert}, {R{\"u}gamer}, {Saggion}, {Saito}, {Salvati}, {Sanchez-Conde},  {Sartori}, {Satalecka}, {Scalzotto}, {Scapin}, {Schmitt}, {Schweizer},  {Shayduk}, {Shinozaki}, {Shore}, {Sidro}, {Sierpowska-Bartosik},  {Sillanp{\"a}{\"a}}, {Sobczynska}, {Spanier}, {Stamerra}, {Stark}, {Takalo},  {Tavecchio}, {Temnikov}, {Tescaro}, {Teshima}, {Tluczykont}, {Torres},  {Turini}, {Vankov}, {Venturini}, {Vitale}, {Wagner}, {Wittek}, {Zabalza},  {Zandanel}, {Zanin}, \& {Zapatero} 2008]{2008Sci...320.1752M}
{MAGIC Collaboration}, {Albert}, {et al.} 2008, Science, 320, 1752

\bibitem[Maraschi et al.(2008)]{2008arXiv0810.0145M} Maraschi, L., 
Foschini, L., Ghisellini, G., Tavecchio, F., 
\& Sambruna, R.~M.\ 2008, arXiv:0810.0145 

\bibitem[{Marconi}, {Risaliti}, {Gilli}, {Hunt},  {Maiolino}, \& {Salvati} 2004]{2004MNRAS.351..169M}
{Marconi}, A., {Risaliti}, G., {Gilli}, R., {Hunt}, L.~K., {Maiolino}, R., \&  {Salvati}, M. 2004, \mnras, 351, 169

\bibitem[Mastichiadis 
\& Moraitis(2008)]{2008A&A...491L..37M} Mastichiadis, A., \& Moraitis, K.\ 2008, \aap, 491, L37 

\bibitem[Merloni et al.(2003)]{2003MNRAS.345.1057M} Merloni, A., Heinz, S., 
\& di Matteo, T.\ 2003, \mnras, 345, 1057 

\bibitem[{M{\"u}cke} \& {Pohl} 2000]{2000MNRAS.312..177M}
{M{\"u}cke}, A. \& {Pohl}, M. 2000, \mnras, 312, 177 (MP00)

\bibitem[Mukherjee 
\& Chiang(1999)]{1999APh....11..213M} Mukherjee, R., \& Chiang, J.\ 1999, Astroparticle Physics, 11, 213 

\bibitem[{Narayan} \& {Quataert} 2005]{2005Sci...307...77N}
{Narayan}, R. \& {Quataert}, E. 2005, Science, 307, 77

\bibitem[{Narumoto} \& {Totani} 2006]{2006ApJ...643...81N}
{Narumoto}, T. \& {Totani}, T. 2006, \apj, 643, 81 (NT06)

\bibitem[{Oda}, {Totani}, \&  {Nagashima} 2005]{2005ApJ...633L..65O}
{Oda}, T., {Totani}, T., \& {Nagashima}, M. 2005, \apjl, 633, L65

\bibitem[Oh(2001)]{2001ApJ...553...25O} Oh, S.~P.\ 2001, \apj, 553, 25 

\bibitem[{Padovani}, {Ghisellini}, {Fabian}, \&  {Celotti} 1993]{1993MNRAS.260L..21P}
{Padovani}, P., {Ghisellini}, G., {Fabian}, A.~C., \& {Celotti}, A. 1993,  \mnras, 260, L21

\bibitem[{Padovani}, {Giommi}, {Landt}, \&  {Perlman} 2007]{2007ApJ...662..182P}
{Padovani}, P., {Giommi}, P., {Landt}, H., \& {Perlman}, E.~S. 2007, \apj, 662,  182

\bibitem[Padovani 
\& Urry(1992)]{1992ApJ...387..449P} Padovani, P., \& Urry, C.~M.\ 1992, \apj, 387, 449 

\bibitem[{Pavlidou} \& {Venters} 2008]{2008ApJ...673..114P}
{Pavlidou}, V. \& {Venters}, T.~M. 2008, \apj, 673, 114

\bibitem[{Pohl} \& {Esposito} 1998]{1998ApJ...507..327P}
{Pohl}, M. \& {Esposito}, J.~A. 1998, \apj, 507, 327

\bibitem[{Protheroe} 1986]{1986MNRAS.221..769P}
{Protheroe}, R.~J. 1986, \mnras, 221, 769

\bibitem[{Salamon} \& {Stecker} 1994]{1994ApJ...430L..21S}
{Salamon}, M.~H. \& {Stecker}, F.~W. 1994, \apjl, 430, L21

\bibitem[{Salamon} \& {Stecker} 1998]{1998ApJ...493..547S}
---. 1998, \apj, 493, 547

\bibitem[{Sazonov}, {Revnivtsev}, {Krivonos},  {Churazov}, \& {Sunyaev} 2007]{2007A&A...462...57S}
{Sazonov}, S., {Revnivtsev}, M., {Krivonos}, R., {Churazov}, E., \& {Sunyaev},  R. 2007, \aap, 462, 57

\bibitem[{Sreekumar}, {Bertsch}, {Dingus},  {Esposito}, {Fichtel}, {Hartman}, {Hunter}, {Kanbach}, {Kniffen}, {Lin},  {Mayer-Hasselwander}, {Michelson}, {von Montigny}, {Muecke}, {Mukherjee},  {Nolan}, {Pohl}, {Reimer}, {Schneid}, {Stacy}, {Stecker}, {Thompson}, \&  {Willis} 1998]{1998ApJ...494..523S}
{Sreekumar}, P., {et al.} 1998, \apj, 494, 523

\bibitem[{Stecker}, {Hunter}, \&  {Kniffen} 2008]{2008APh....29...25S}
{Stecker}, F.~W., {Hunter}, S.~D., \& {Kniffen}, D.~A. 2008, Astroparticle  Physics, 29, 25

\bibitem[{Stecker}, {Malkan}, \&  {Scully} 2006]{2006ApJ...648..774S}
{Stecker}, F.~W., {Malkan}, M.~A., \& {Scully}, S.~T. 2006, \apj, 648, 774

\bibitem[{Stecker} \& {Salamon} 1996]{1996ApJ...464..600S}
{Stecker}, F.~W. \& {Salamon}, M.~H. 1996, \apj, 464, 600 (SS96)

\bibitem[Stecker \& Salamon(2001)]{2001AIPC..587..432S} Stecker, F.~W., \& Salamon, M.~H.\ 2001, Gamma 2001: Gamma-Ray Astrophysics, 587, 432 

\bibitem[{Stecker}, {Salamon}, \&  {Malkan} 1993]{1993ApJ...410L..71S}
{Stecker}, F.~W., {Salamon}, M.~H., \& {Malkan}, M.~A. 1993, \apjl, 410, L71

\bibitem[{Strong}, {Moskalenko}, \&  {Reimer} 2004a]{2004ApJ...613..956S}
{Strong}, A.~W., {Moskalenko}, I.~V., \& {Reimer}, O. 2004a, \apj,  613, 956

\bibitem[{Strong}, {Moskalenko}, \&  {Reimer} 2004b]{2004ApJ...613..962S}
---. 2004b, \apj, 613, 962

\bibitem[Taylor 
\& Silk(2003)]{2003MNRAS.339..505T} Taylor, J.~E., \& Silk, J.\ 2003, \mnras, 339, 505 

\bibitem[{Thompson} \& {Fichtel} 1982]{1982A&A...109..352T}
{Thompson}, D.~J. \& {Fichtel}, C.~E. 1982, \aap, 109, 352

\bibitem[{Totani} \& {Takeuchi} 2002]{2002ApJ...570..470T}
{Totani}, T. \& {Takeuchi}, T.~T. 2002, \apj, 570, 470

\bibitem[Totani(2006)]{2006PASJ...58..965T} Totani, T.\ 2006, \pasj, 58, 
965 

\bibitem[{Ueda}, {Akiyama}, {Ohta}, \&  {Miyaji} 2003]{2003ApJ...598..886U}
{Ueda}, Y., {Akiyama}, M., {Ohta}, K., \& {Miyaji}, T. 2003, \apj, 598, 886 (U03)

\bibitem[Urry et al.(1991)]{1991ApJ...382..501U} Urry, C.~M., Padovani, P., 
\& Stickel, M.\ 1991, \apj, 382, 501 

\bibitem[Urry 
\& Padovani(1995)]{1995PASP..107..803U} Urry, C.~M., \& Padovani, P.\ 1995, \pasp, 107, 803 

\bibitem[Venters 
\& Pavlidou(2007)]{2007ApJ...666..128V} Venters, T.~M., \& Pavlidou, V.\ 2007, \apj, 666, 128 

\bibitem[{Watanabe}, {Hartmann}, {Leising}, \&  {The} 1999]{1999ApJ...516..285W}
{Watanabe}, K., {Hartmann}, D.~H., {Leising}, M.~D., \& {The}, L.-S. 1999,  \apj, 516, 285

\end{thebibliography}

\appendix 
\section{The Blazar SED Sequence Formulations}
We define $\psi(x) \equiv \log_{10} [\nu L_\nu / (\rm erg \ s^{-1})]$ with
$x \equiv \log_{10} (\nu / \rm Hz)$ ($\nu$ in rest-frame).
The empirical SED sequence model of blazars
is the sum of the synchrotron $[\psi_s(x)]$ and IC $[\psi_c(x)]$
emissions, i.e.,
\begin{eqnarray}
\psi(x) = \log_{10} [10^{\psi_s(x)} + 10^{\psi_c(x)} ] \ .
\end{eqnarray}
Each of the two components is described by a combination of a linear
and a parabolic functions at low and high photon frequencies, and we
define $x_{\rm tr, s}$ and $x_{\rm tr, c}$ as the linear-parabolic
transition frequencies for the synchrotron and IC components,
respectively.  We take $\psi_R \equiv \log_{10} [L_R / (\rm erg \ s^{-1})]$
as a reference of a blazar luminosity, where $L_R$ is $\nu L_\nu$
luminosity in the radio band ($\nu_R = $5 GHz or $x_R = 9.698$). The
linear part of the synchrotron component is described as follows:
\begin{eqnarray}
\psi_s(x) = \psi_{s1}(x) \equiv  (1 - \alpha_s) (x - x_R)+\psi_R 
\quad (x < x_{\rm tr, s}) \ ,
\end{eqnarray}
where 
$\alpha_s = 0.2$ is the energy flux index (i.e., $L_\nu \propto
\nu^{-\alpha_s}$).  The parabolic part of the synchrotron component is
characterized by the $\nu L_\nu$ peak frequency $\nu_s$ (or
corresponding $x_s$), as
\begin{eqnarray}
\psi_s(x) = \psi_{s2}(x) \equiv - [(x-x_s)/\sigma]^2 + \psi_{s, p} 
\quad (x \geq x_{\rm tr, s}) \ ,
\end{eqnarray}
where $\sigma$ is a parameter that controls the width of the parabolic
function.  The peak luminosity $\psi_{s, p}$ of the synchrotron
component is determined if $x_{\rm tr, s}$, $x_s$, and $\sigma$ are
given, by requiring the continuity between the linear and parabolic
parts, i.e., $\psi_{s1}(x_{\rm tr, s}) = \psi_{s2}(x_{\rm tr, s})$.
The result is:
\begin{equation}
\psi_{s,p}= (1-\alpha_s)(x_{\rm tr, s} - x_R)
+ \psi_R + \left(\frac{x_{\rm tr, s}
 - x_s }{ \sigma}\right)^2 \ .
\end{equation}

The linear part of the IC component (roughly in the hard X-ray band)
is defined as:
\begin{eqnarray}
\psi_c(x) = \psi_{c1}(x) \equiv (1-\alpha_c)(x - x_X) + \psi_X
\quad (x < x_{\rm tr, c}) \ ,
\end{eqnarray}
where 
the power index is $\alpha_c = 0.6$ (different from the
synchrotron component), and $\psi_X$ is the luminosity of
IC component at the reference frequency 
$\nu_X = 2.42 \times 10^{17}$ Hz = 1 keV$/h_p$, or $x_X = 17.383$. 
The parabolic part of the IC component is characterized by the
$\nu L_\nu$ peak frequency $\nu_c$ (or corresponding to $x_c$) with
the same width parameter $\sigma$ as the synchrotron component, i.e.,
\begin{eqnarray}
\psi_c(x) = \psi_{c2}(x) \equiv
 - [(x-x_c)/\sigma]^2 +\psi_{c, p}	
\quad (x \geq x_{\rm tr, c}) \ ,
\end{eqnarray}
where $\psi_{c,p}$ is the peak luminosity of
the IC component.  The linear-parabolic transition frequency 
$x_{\rm tr, c}$ is determined
by requiring continuity, $\psi_{c1}(x_{\rm tr, c}) = \psi_{c2}(x_{\rm tr, c})$
 and the result is:
\begin{equation}
x_{\rm tr, c}= \frac{-\zeta-\sqrt{\zeta^2 - 4\eta}}{2} \ ,
\end{equation}
where
\begin{eqnarray}
\zeta &=& \sigma^2 (1-\alpha_c) - 2 x_c \ ,\\
\eta &=& x_c^2 + \sigma^2 [\psi_X - x_X (1-\alpha_c) - \psi_{c,p}] \ .
\end{eqnarray}

Then, the SED sequence is determined when $x_{\rm tr, s}$, $x_s$,
$\sigma$, $\psi_X$, $x_c$, and $\psi_{c, p}$ are given as functions of
$\psi_R$.  First we set $x_{\rm tr, s}$=$\log_{10}$($5\times 10^{10}$) and
$10^{x_c - x_s} = \nu_c/\nu_s =5\times10^8$, for all luminosity range
of blazars. To get a good fit in all luminosity range, we divide the
luminosity into two ranges of $\psi_R \leq 43$ and $\psi_R > 43$.  The
parameters relevant for the synchrotron component in the $\psi_R
\leq$ 43 range are determined as follows.  The synchrotron peak
frequency is determined by
\begin{equation}
x_s = - \xi (\psi_R -43) + 14.47 \ ,
\end{equation}
with $\xi = 0.88$.
The width parameter $\sigma$ is determined by the following scaling
law:
\begin{equation}
\sigma=0.0891 \, x_{\rm s} + 1.78 \ .  
\end{equation}
In this case, the connection at $x=x_{\rm tr, s}$ is not smooth (i.e.,
not differentiable), but the discontinuity of the derivative is not
significant.  
In the range of $\psi_R >$ 43, we change the synchrotron peak
frequency $x_s$ by adopting $\xi = 0.4$, and the width parameter
$\sigma$ is determined so that the connection at $x_{\rm tr, s}$
is differentiable, i.e., 
\begin{equation}
\sigma=[2(x_{\rm s}-x_{\rm tr, s})/(1 - \alpha_s)]^{1/2}  \ .
\end{equation}

For the IC component parameters, 
the IC peak luminosity $\psi_{c,p}$ is assumed to be the
same as the synchrotron peak, i.e., $\psi_{c,p} = \psi_{s,p}$,
at $\psi_R \leq 43$, while it is 
determined by the
following formula at $\psi_R > 43$:
\begin{eqnarray}
\psi_{c,p} = \beta (\psi_R -43)^\delta + \epsilon ,
\end{eqnarray}
where $\beta = 1.77$, $\delta = 0.718$, and $\epsilon = 45.3$. 
Still, the normalization of the IC linear part,
$\psi_X$, remains to be determined. To get a good fit to
the X-ray data, we define the dependence of this parameter on
$\psi_R$ as follows:
\begin{eqnarray}
\psi_X = \left\{ 
    \begin{array}{rl}
      (\psi_R -43 ) + \psi_{X, 43}       & \quad (\psi_R \leq 43) \\
      1.40 \, (\psi_R - 43) + \psi_{X, 43}  
                                 & \quad (43 < \psi_R \leq 46.68) \\
      1.40 \, (46.68 - 43) + \psi_{X, 43}       & \quad (46.68 < \psi_R) \\
    \end{array}
\right.
\end{eqnarray}
with $\psi_{X, 43} = 43.17$. The parameter $\psi_X$ is kept constant
at $\psi_R > 46.68$, because $\psi_{c1}(x) = \psi_{c2}(x)$ does not
have solution and hence $x_{\rm tr, c}$ cannot be obtained, when we
extrapolate the $\psi_X$ formula at $\psi_R \leq 46.68$ into $\psi_R >
46.68$.  At $\psi_R = 46.68$, $\psi_{c, p}$ becomes 49.81, which is
brighter than the maximum $L_\gamma$ of the EGRET blazars, and is
larger than the characteristic $L_\gamma^*$ corresponding to the break
X-ray luminosity $L_X^*$ in the AGN XLF by a factor of $\sim
60$. Therefore, the treatment of $\psi_X$ at such large luminosity
range is not important when we consider the overall properties of
blazars and EGRB.  The linear-parabolic connection of the IC component
at $x_{\rm tr, c}$ is not as smooth as that of the synchrotron
component at $x_{\rm tr, s}$, but this is inevitable to fit the SED in
the X-ray region.

It should be noted that the formulations of $x_s$, $\sigma$, $\psi_{c,
  p}$, and $\psi_X$ are continuous at $\psi_R = 43$ and
46.68. Therefore the SED in this formula always changes continuously
with $\psi_R$ in the whole luminosity range, which was the motivation
of constructing this new formulation.

\end{document}